# Electromechanical System Design for Self-Balancing Robot

Md. Abid Al Morshed, Md. Mustakim Hayder, Tayfur Rahman Maruf

Department of Mechanical Engineering
Bangladesh University of Engineering and Technology

## Abstract:

Self-balancing robot is based on the principle of Inverted pendulum, which is a two-wheel vehicle balances itself up in the vertical position with reference to the ground. It consists of both hardware and software implementation. Mechanical model based on the state space design of the cart, pendulum system. To find its stable inverted position, we used a generic feedback controller (i.e., PID controller). According to the situation we have to control both angel of pendulum and position of cart. Mechanical design consists of two DC gear motor with encoder, one Arduino Micro-controller, IMU (inertial mass unit) sensor and motor driver as a basic need. IMU sensor which consists of accelerometer and gyroscope gives the reference acceleration and angle with respect to ground (vertical), When encoder which is attached with the motor gives the speed of the motor. These parameters are taken as the system parameter and determine the external force needed to balance the robot up. It will be prevented from falling by giving acceleration to the wheels according to its inclination from the vertical. If the bot tilts by an angle, then in the frame of the wheels; the center of mass of the bot will experience a pseudo force which will apply a torque opposite to the direction of tilt. We used a HC05 Bluetooth module to control the robot via our smartphone.

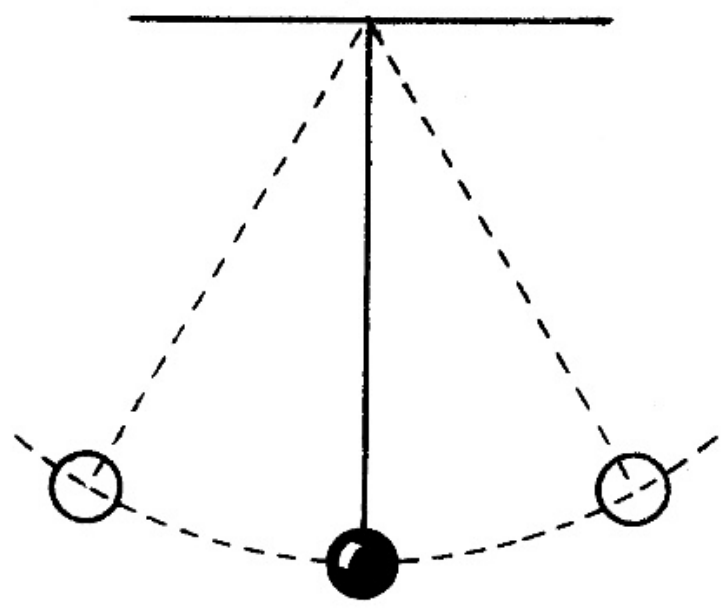

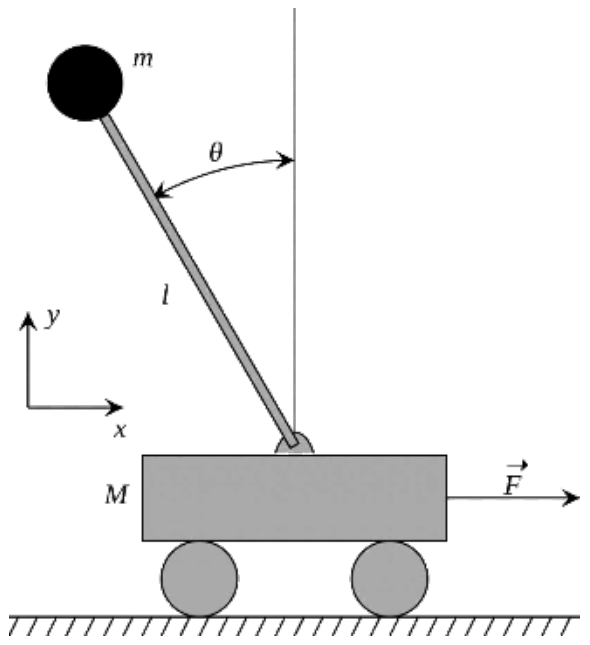

# 1. Introduction:

To make a self-balancing robot, it is essential to solve the inverted pendulum problem or an inverted pendulum on cart. While the calculation and expressions are very complex, the goal is quite simple: the goal of the project is to adjust the wheels' position so that the inclination angle remains stable. Inherently self-balancing robot is unstable, and it would roll around the wheels' rotation axis without external control and eventually fall. When the robot starts to fall in one direction, the wheels should move in the inclined direction with a speed proportional to angle and acceleration of falling to correct the inclination angle.

The robot returns to the right position if motor driving occurs in right direction. The robot is naturally unstable although, it has many favors over the statically stable multi wheeled robots. A special electromechanical system in which the robot has to be based on balances itself onto a pair of wheels while standing tall. If the base on which the robot stands is not stable or the platform is not balanced, the robot tends to be falling off from the vertical axis. This time a gyro chip is needed to provide the PID controller about the angular position of the base of the self-balancing robot.

A self-balancing algorithm is programmed into the controller and the controller drives the motors either clockwise or anticlockwise to balance the basement by a pulse width modulation (PWM) control signal. The robot has to be worked upon any type of surface based on two stepper motors constructed with wheel one for each.

### 1.1 Motivation:

The primary incentive of the project is to develop a general understanding of control theory. For the last few decades, "the inverted pendulum has been the most popular benchmark, among others, for teaching and research in control theory and robotics. Furthermore, the material and methods learnt have a wide array of applications; for example, inverted pendulums have been used to model human locomotion, which then was used to develop bipedal robots. Besides learning about the theoretical aspects, the project also incorporated a practical side. These are a wide array of important skills applicable to many tasks, in future projects, as an engineer.

### 1.2 Basic Aim:

- To demonstrate the methods and techniques involved in balancing an unstable robotic platform on two wheels.
- To obtain a dynamic model and working prototype
- To stabilize the unstable system by maintaining the upright position.
- To obtain minimum settling time.
- To design a complete digital control system with the state space model that will provide the needed to control the developed system remotely.

### 1.3 Scope:

- To build a demonstrator and implement a code that enables the demonstrator to balance and to be controlled over a wireless connection.
- To enable the robot to carry packages without losing balance.
- To be used as autonomous trolleys in malls, hospitals and airports
- To be used for smart gardening purposes.

Perhaps the most famous real-world example of a self-balancing robot is the **Segway**, a two-wheeled motorized personal vehicle.

**1.4 Approach used in carrying out the project:**

- Software modeling: We used Solid works and MATLAB to design a draft model of our project. We used fritzing to design our circuit.
- Purchasing Parts: We ordered our parts from two online shops, which are Robotics BD and Speedy Tech.
- Building Frame: We used a hardboard to build the frame.
- Circuit Design: Soldering process was involved. We used the diagram we got from fritzing to build the circuit.
- Code: We used GitHub for our source code. We changed it according to our requirements.
- Uploading and Final Result: It is the last part of our whole process. Arduino IDE was used to upload the code. To control the **robot,** we used EZ GUI app from eziosoft.

# 2. Background or Literature Review:

Control systems reduce the involvement of humane supervision in many aspects of our life. By changing the input of a system, the controller automatically tries to match the desired output. Self-balancing robots are a noteworthy example of implementation of control systems. The desired outcome here is for the robot to maintain zero angle with the vertical. With the help of a control system the robot monitors the tilt and accelerates accordingly to maintain vertical position without any human command.

Using a two-wheel locomotion system in a robot can be seen as mimicking man walking on his feel. This is both interesting and has advantages over four wheeled locomotion. In an idle state the robot will take up less space, will be able to maneuver in congested areas and by modifying the system it can be made to get after falling down without much hassle. There is a lot of potential in using self-balancing systems in a robot's locomotion.

This project will allow us to demonstrate the usefulness of a self-balancing robot while learning to develop a control system.

# 3. Design and Methodology

**3.1 Design Goals and requirements:**

The focus of the project was to build a demonstrator and implement a code that enables the demonstrator to balance and to be controlled over a wireless connection.

Inherently self-balancing robot is unstable, and it would roll around the wheels' rotation axis without external control and eventually fall. A special electromechanical system in which the robot has to be based on, balances itself onto a pair of wheels while standing tall. The robot has to work on any type of surface based on two motors with wheels attached.

### 3.2 Design Process:

The self-balancing robot gets balanced on a pair of wheels having the required grip providing sufficient
friction. For maintaining vertical axis two things must be done, one is measuring the inclination angle and other is controlling of motors to move forward or backwards to maintain 0°angle with vertical axis. For measuring the angle, two sensors, accelerometer and gyroscope are used. Accelerometer can sense either static or dynamic forces of acceleration and Gyroscope measures the angular velocity. The outputs of the sensors are fused using a Complementary filter. Sensors measure the process output say α which gets subtracted from the reference set-point value to produce an error. Error is then fed into the PID where the error gets managed in three ways. After the PID algorithm processes the error, the controller produces a control signal μ. PID control signal then gets fed into the process under control. Process under PID control is two wheeled robot. PID control signal will try to drive the process to the desired set-point value that is 0° in vertical position by driving the motors in such a way that the robot is balanced.

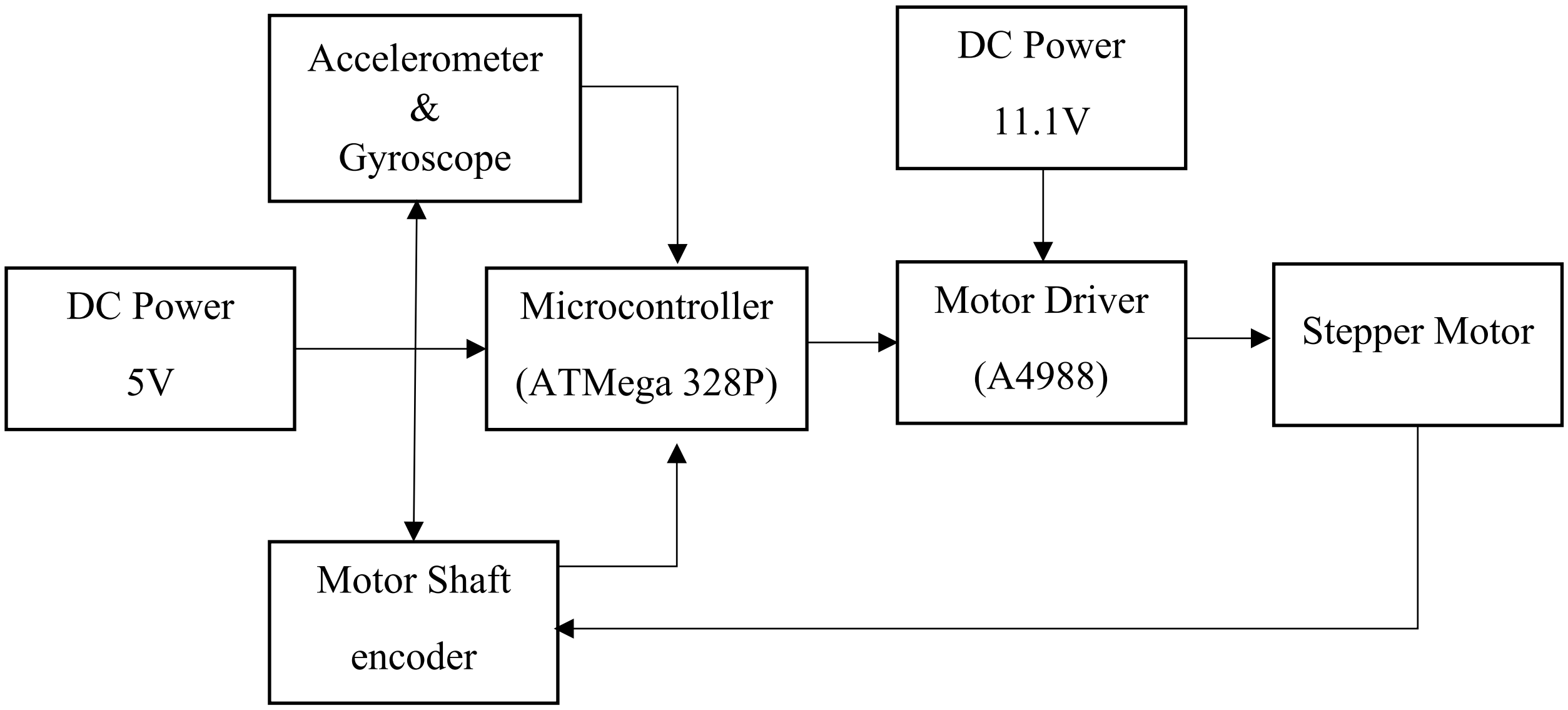

Fig.1: Block diagram of self-balancing robot

In order to design a control system, a mechanical model of the robot that describes the movement of the robot and the forces that act on the pendulum and the wheels was derived. The system was modeled as cart and a pendulum, see figure 2. The system was simplified by only studying the forces and angles in the x- and y-axis. When deriving the mechanical system, it was assumed that the wheels never lose contact with the ground, resulting in the wheels not having any movement in the y-axis. Any disturbances due to the change of the center of gravity was neglected.

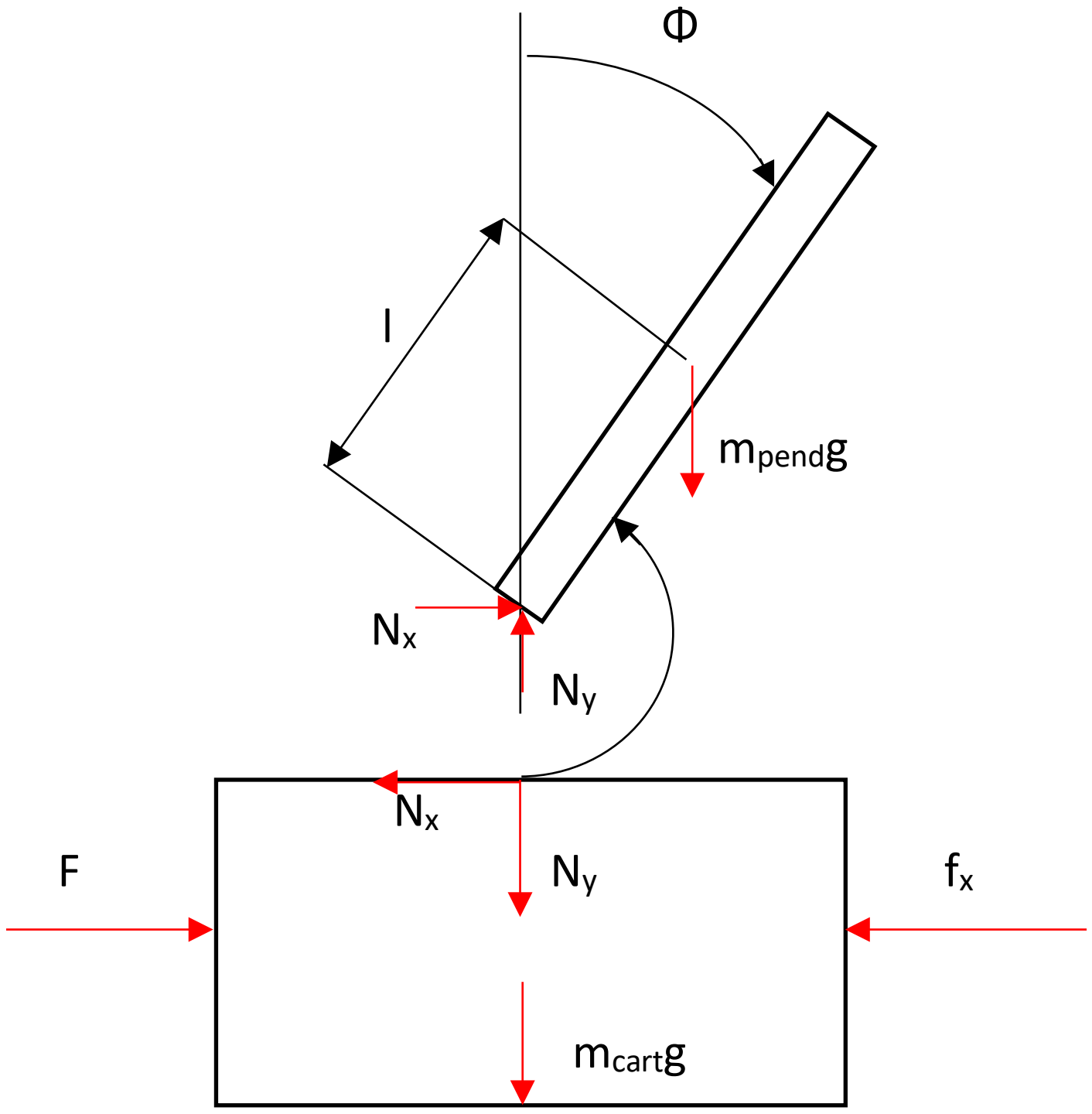

Fig. 2. Figure showing the forces acting on the system.

**3.3 Control system design:**

It is a challenging work to balance an inverted pendulum, because naturally it is unstable. The small error or disturbance from equilibrium position takes away from equilibrium forcefully that further destabilizes the system. So, keeping balance at a slightly non-equilibrium requires precise control to immediately correct any errors in tilt the instant they happen [1]. To deal with this problem, a PID controller is employed that uses tilt feedback to control the torque of the motors and keep the robot balanced. A PID controller continuously measures a process variable and calculates an error value (angle from the vertical), which is the deviation of the process variable from some desired value (0 degrees from the vertical)[2]. The PID controller try to minimize this type of error over time by continuously adjusting a control variable (motor torque) according to the following equation, where *u(t)* is the control variable, *e(t)* is the current error in the process variable, and Kp, Ki, and Kd are coefficients that must be tuned to achieve the desired behavior of the controller:

$$u(t) = K_p e(t) + K_i \int_0^t e(t)dt + K_d \frac{de(t)}{dt}$$

**3.4 Data acquisition:**

Both the accelerometer and gyroscope data are used to obtain the angular position of the object. Gyroscope does this by integrating the angular velocity over time. From the position of the gravity vector (g-force) angular position is obtained using the accelerometer. In both these cases, it is very hard to use without a filter. Accelerometer measures all forces that are working on an object, as well as it will also see a lot more than just the gravity vector. Every small force working on the object disturbs measurements in a great amount. While working on an actuated system the forces that drive the system will also be visible on the sensor as well. The accelerometer data is reliable in case of a long term, for that reason a "low pass" filter is required. In the case of Gyro, it is very easy to obtain accurate data which is not susceptible to external forces. But because of the integration over time, the measurement has a tendency to drift, not returning to level zero when the system went back to its original position. This data is reliable only for a short term; it shows drift on the long term. The complementary filter provides data both for short and long term from gyroscope and accelerometer respectively. The filter is shown in Figure 3.

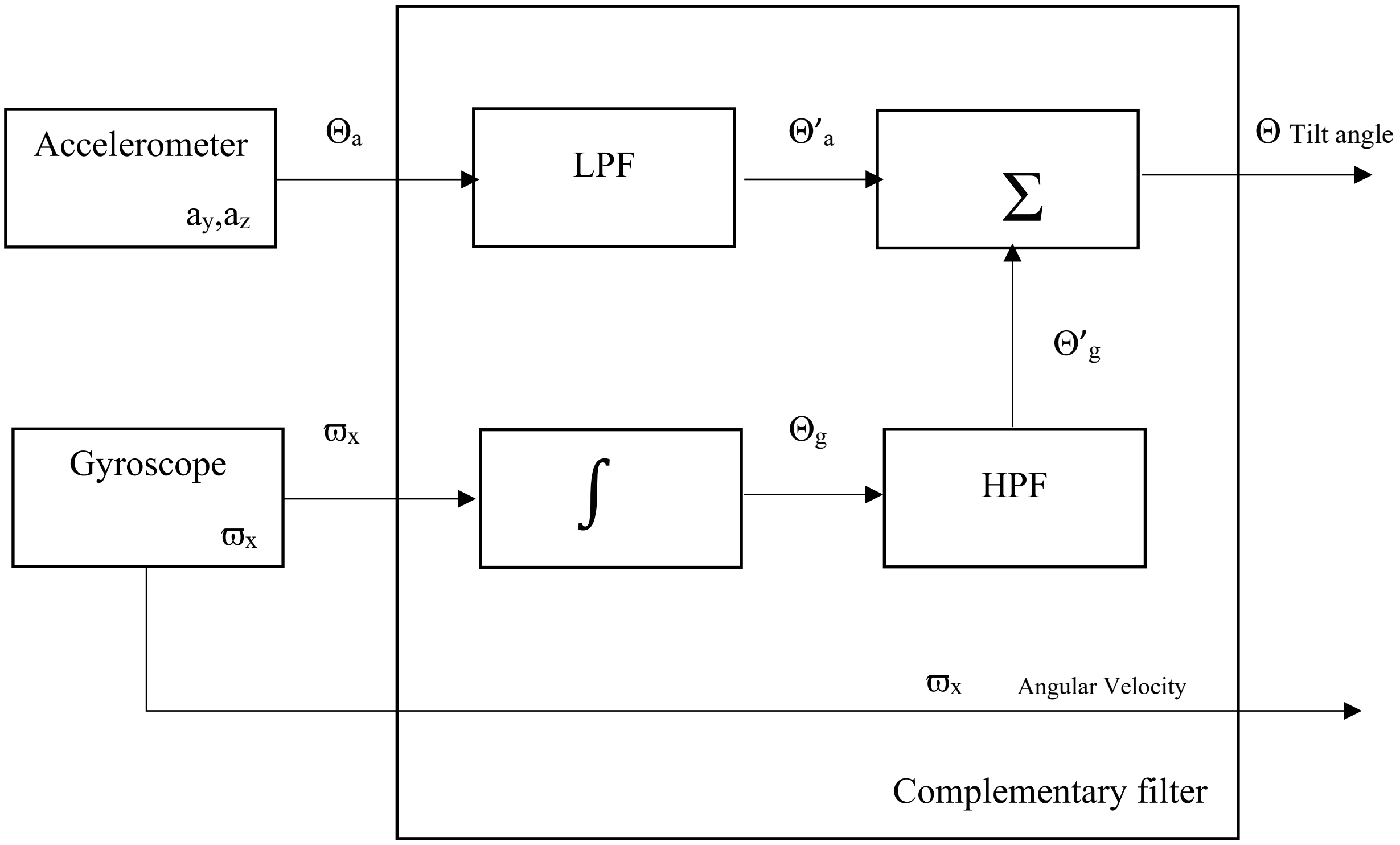

Fig.3. Complementary Filter Block

Arduino nano and MPU 6050 are used to acquire data and filter. Basic connection diagram is shown in Figure 4

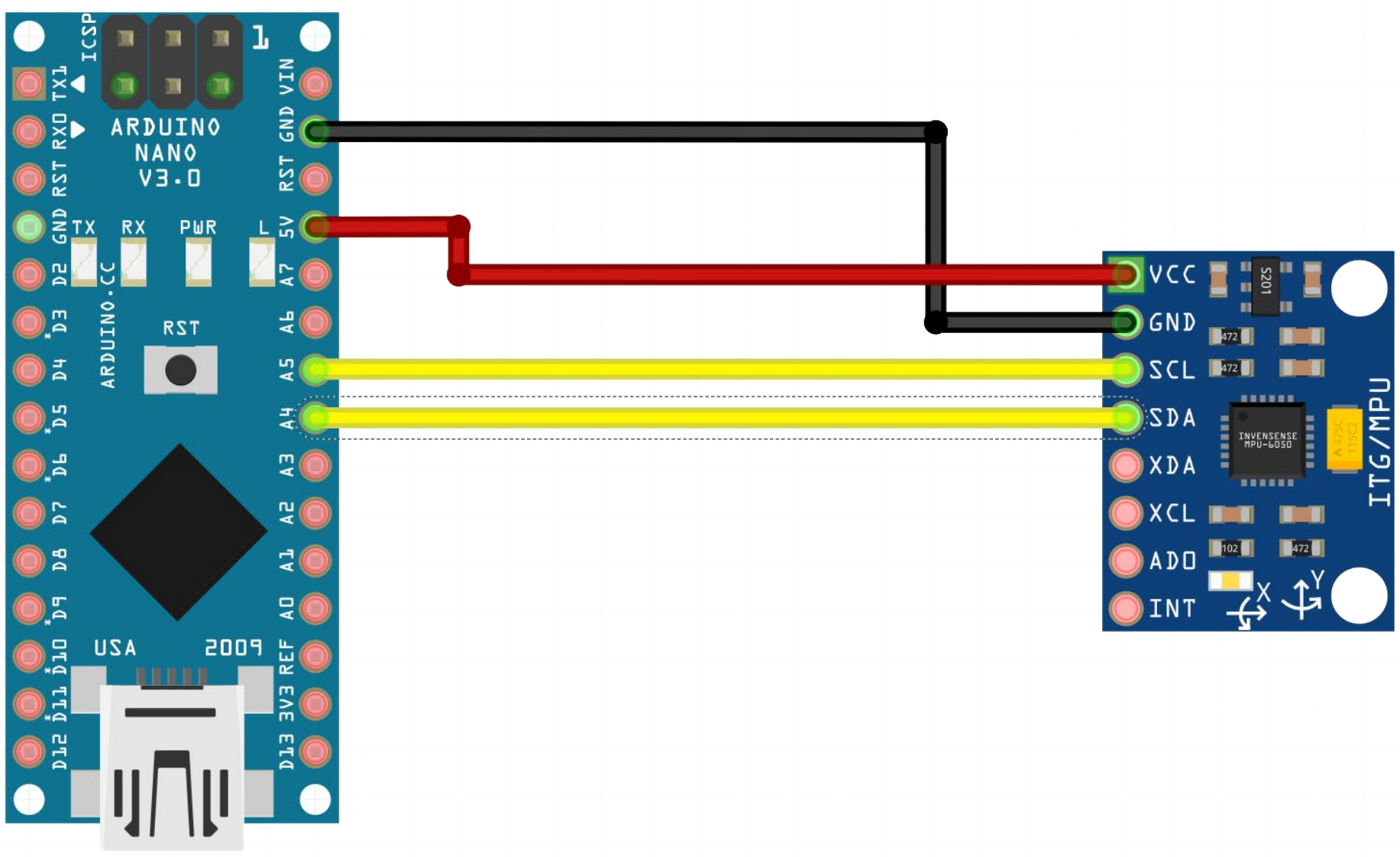

Fig.4. Arduino Nano and MPU 6050 connection

The control system of the two wheels balancing robot in this project is illustrated by the block diagram in Figure 5.

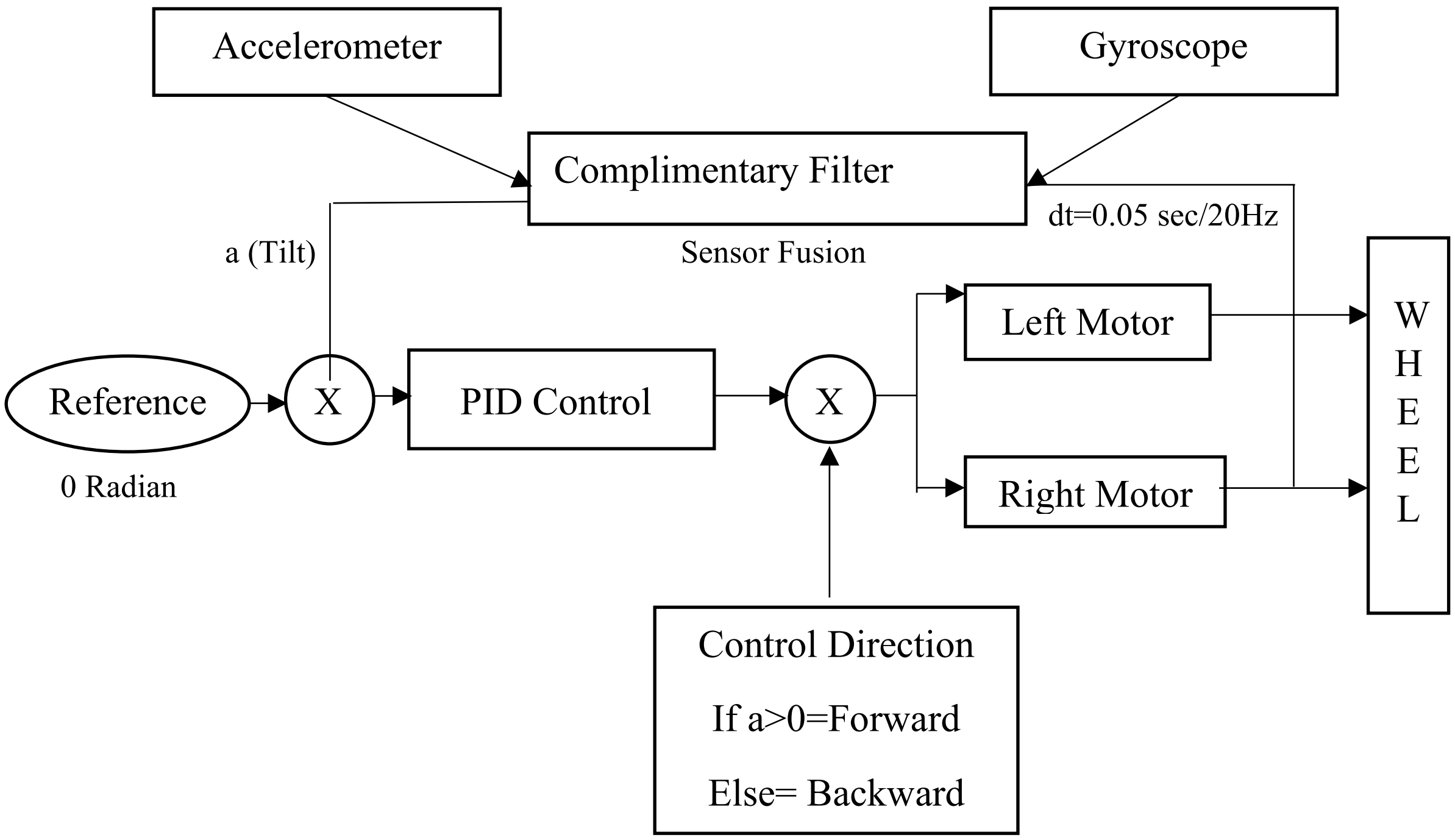

Fig.5. control system of the two wheels balancing robot

The robot control software is inside the microcontroller module and the program flowchart is shown in Figure below:

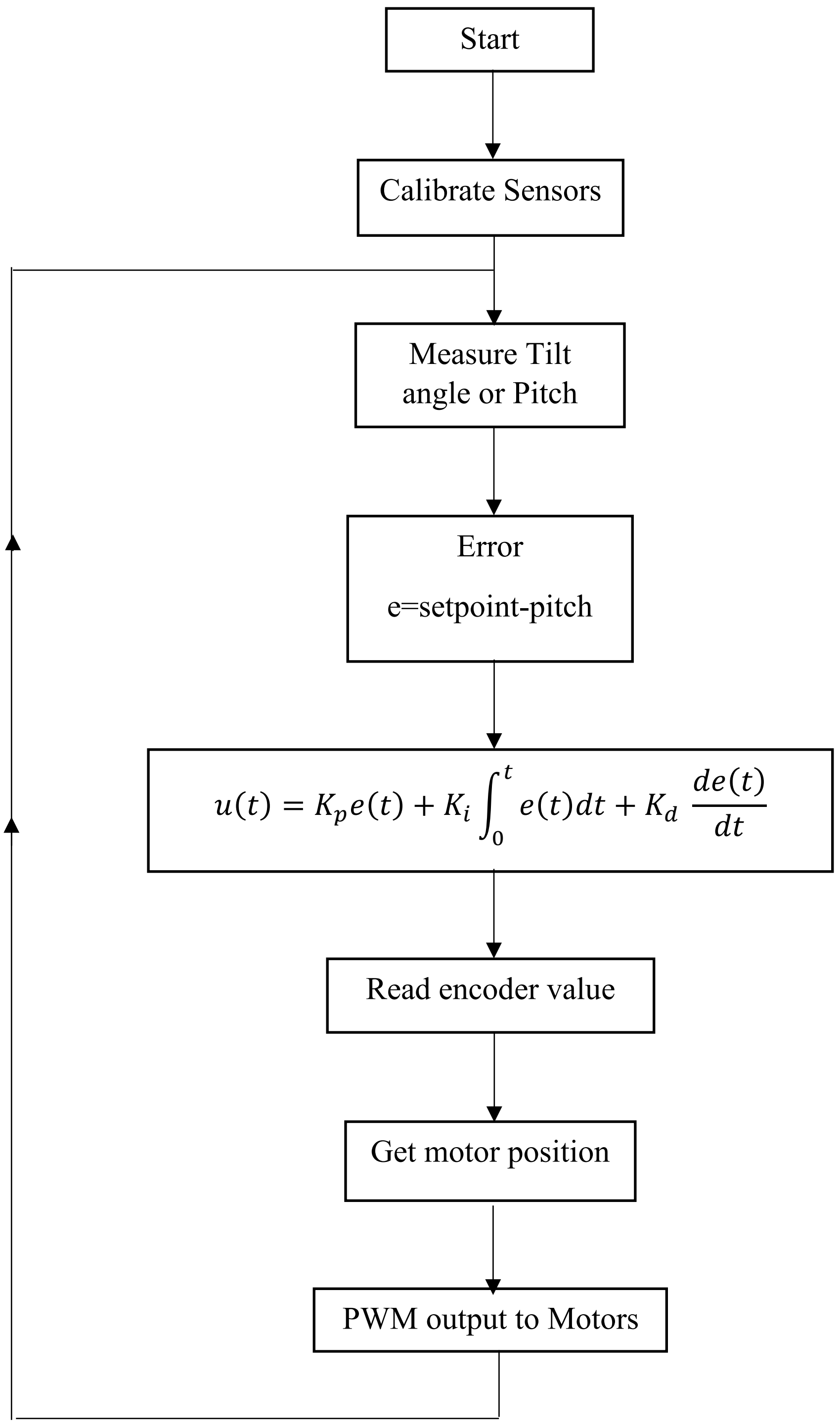

The speed of the motor can be adjusted by the PWM by adjusting the duty cycle and hence the voltage supplied to the motor. Using the PWM method saves the cost of acquiring a Digital to Analogue Converter. Another benefit of using the PWM is that the signal remains digital, and no digital-to-analogue conversion is necessary, by doing so the noise effects are minimized. Changing the set point for the PID controller individually for two motors can control the translational motion of the robot. The motor power increases by the proportional term as the system leans further over and decreases the motor power as the system approaches the upright position [3] [4]. A gain factor, Kp, determines how much power to apply to the motor for any given lean, as follows:

$$Proportional\ Term = K_p * Error$$

The differential term of the PID algorithm acts as a damper reducing oscillation. Another gain factor, $K_d$ determines how much power is applied to the motor according to the following equation:

$$Differential\ Term = K_d * (Error - Last\ Error)$$

Finally, neither the proportional nor differential terms of the algorithm will remove all of the lean because both terms go to zero as the orientation of the system settles near vertical. The integral term sums the accumulated error tries to drive the lean to zero as follows:

$$Output\ integral\ Term = K_i * Sum\ of\ Error$$

The output of the PID controller is:

$$PWM = Proportional\ Term + Integral\ Term + Differential\ Term$$

The output of the Motor PWM as above will be used as the set-point for the motor.

$$Error\ in\ motor\ speed = Motor\ set\ point - Current\ speed\ reading\ of\ the\ motor$$

$$Output\ proportional\ term\ of\ motor\ speed = K_p * Error\ in\ Motor\ Speed$$

$$Output\ differential\ term\ of\ motor\ speed = K_d * (Error\ in\ motor\ speed - Last\ error\ in\ motor\ speed)$$

$$Output\ integral\ term\ of\ motor\ speed = Sum\ of\ Error\ of\ motor\ speed$$

Finally,

$$The\ motor\ speed = proportional\ term\ of\ motor\ speed + differential\ term\ of\ motor\ speed + integral\ term\ of\ motor\ speed$$

For tuning the PID control of motor speed, the value of $K_p$, $K_i$ and $K_d$ is get by trial-and-error method.

**3.5 Implementation of wireless control:**
To steer the robot remotely Bluetooth communication will be used, a radio based wireless network. To obtain a Bluetooth connection between two devices the receiver (HC-05) acts as a access point that the other device (smart phone) can connect to. The other device acts as a station that connects to the access point. Latency is the time that passes between a user action at the station and the response at the access point, a higher latency leads to a longer delay from when the user sends a steering command to when the command is executed by the robot. One of the principle causes of higher latency is the distance between the access point and station and baud rate.

To enable forward and backward movement the angle set point of the PID controller will be increased or decreased, resulting in the robot leaning forward or backward in order to maintain the angle set point. This will make the robot move in the leaning direction.

To enable turns the output signals to the motors are divided for the left motor and the right motor. By reducing the pulse frequency to one motor and increasing the frequency to the other motor it will enable the robot to turn.

**3.6 Draft model of the bot:**
We used SolidWorks to build a draft model of our robot

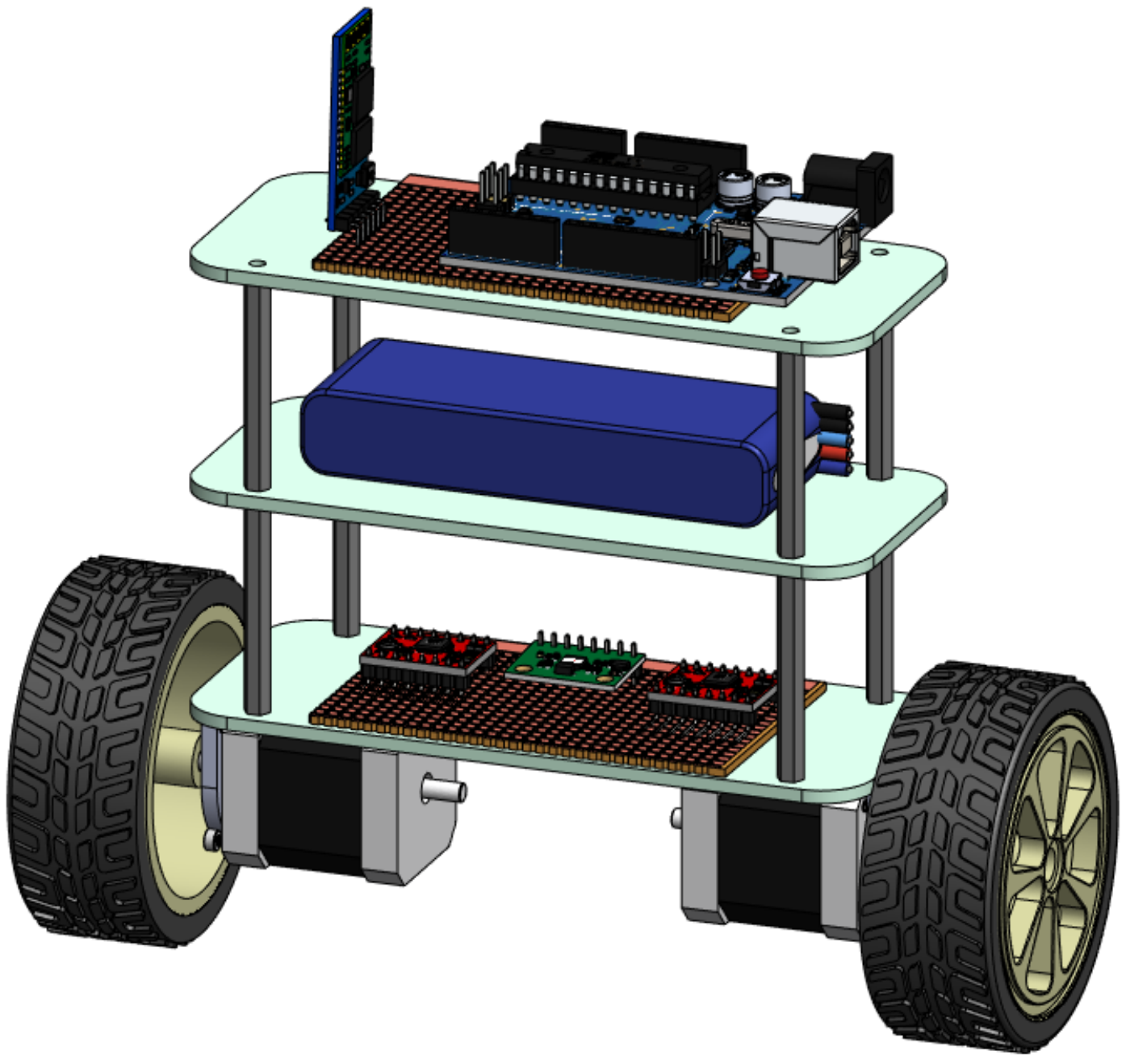

Also, we used Simulink to model the control system.

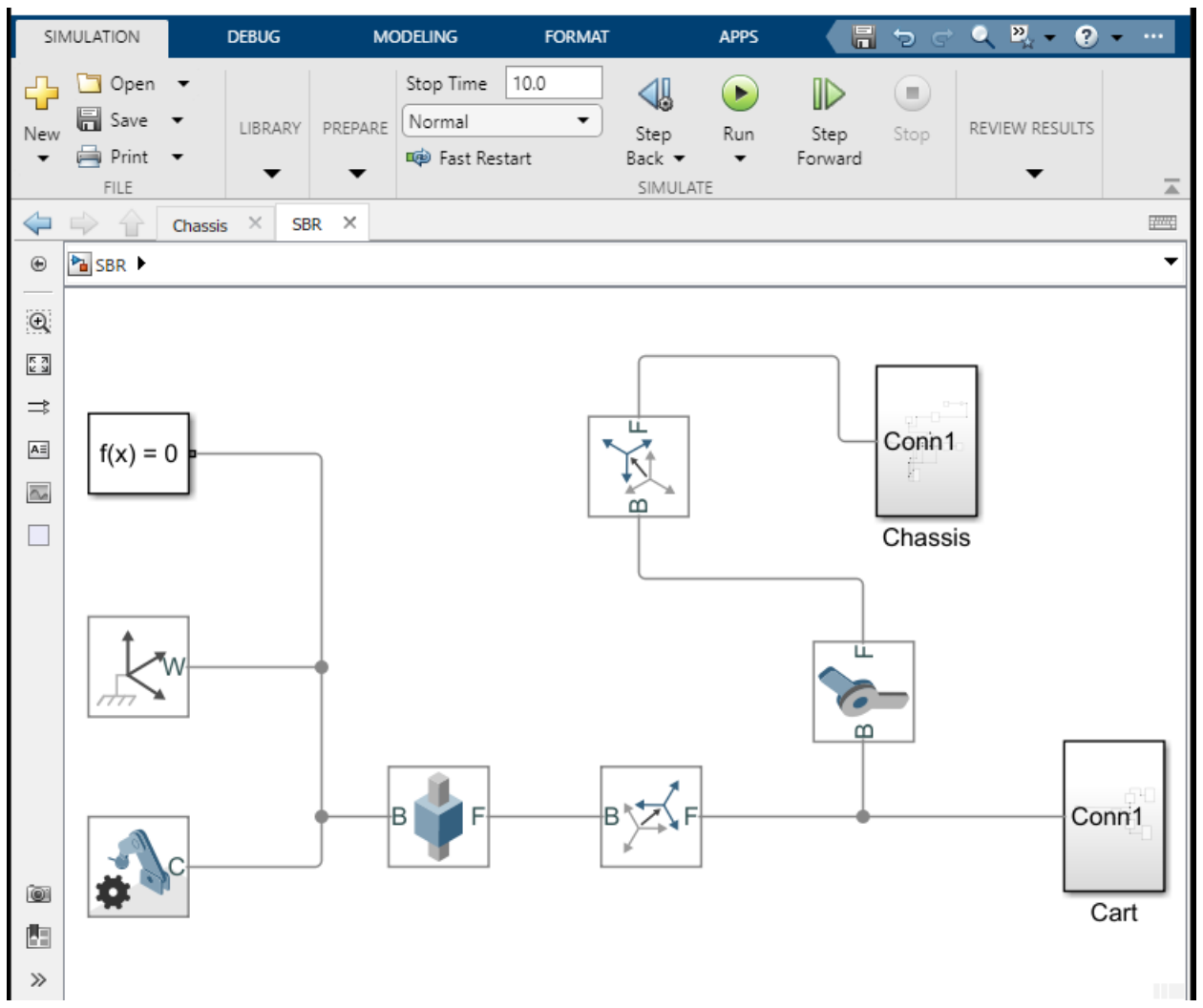

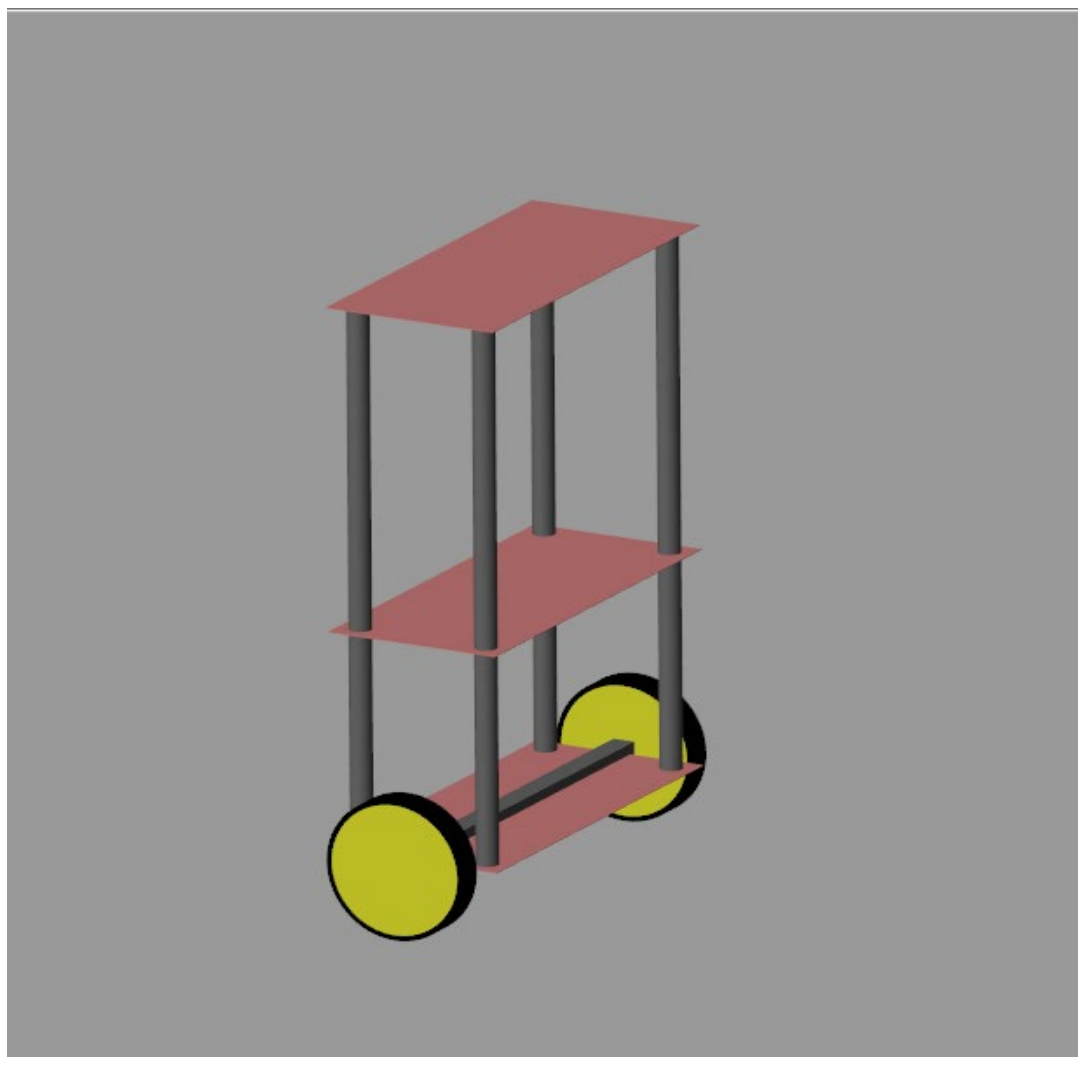

To make our circuit diagram and model a pcb we used fritzing.

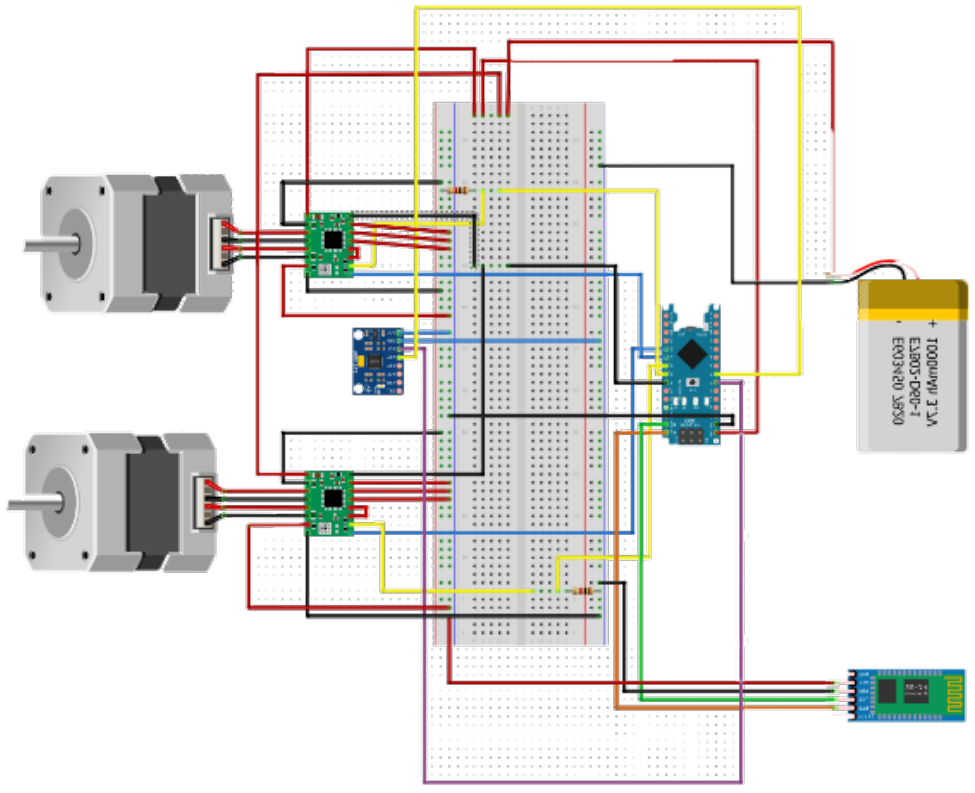

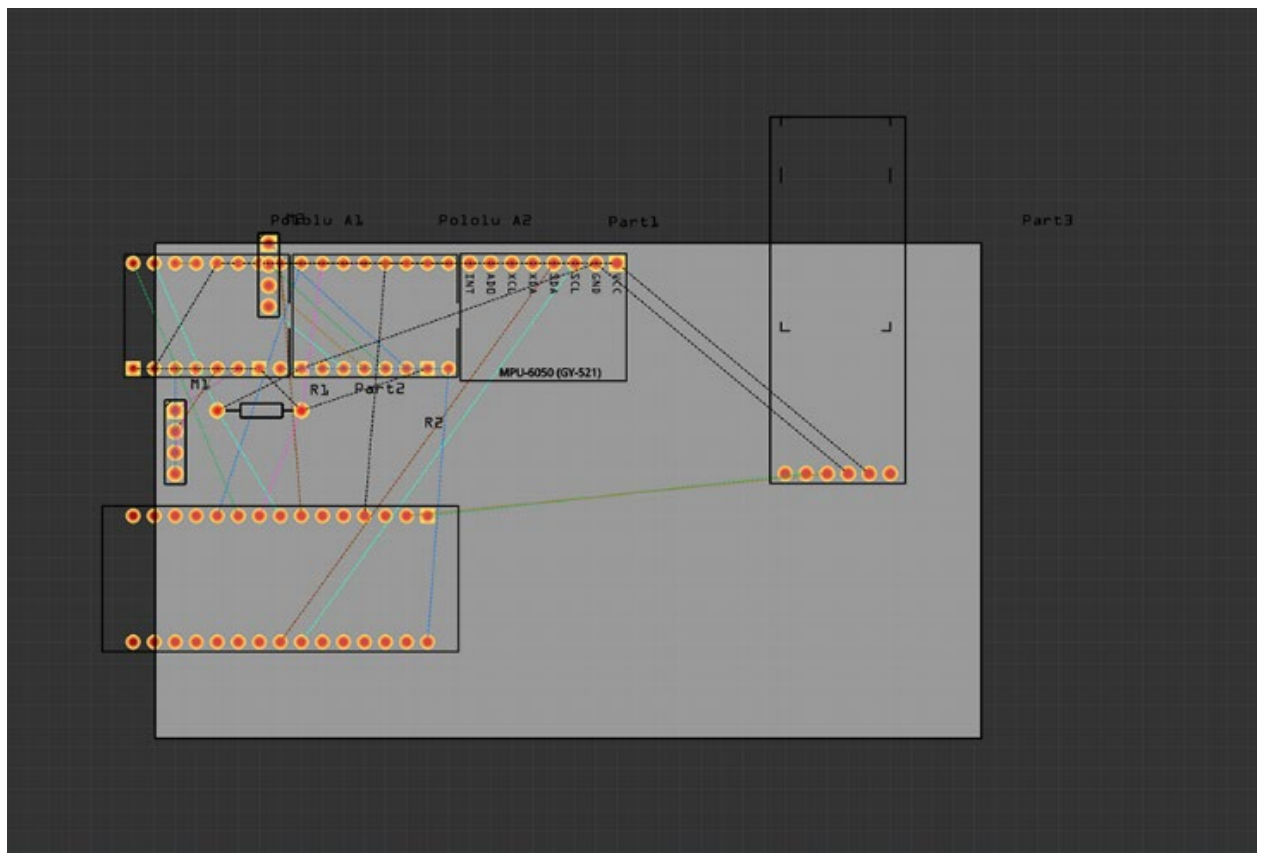

# 4. Components Used:

### 4.1 Mechanical:

1. Hardboard body: The dimensions of the platforms were determined by adding allowance to the Veroboard dimension.

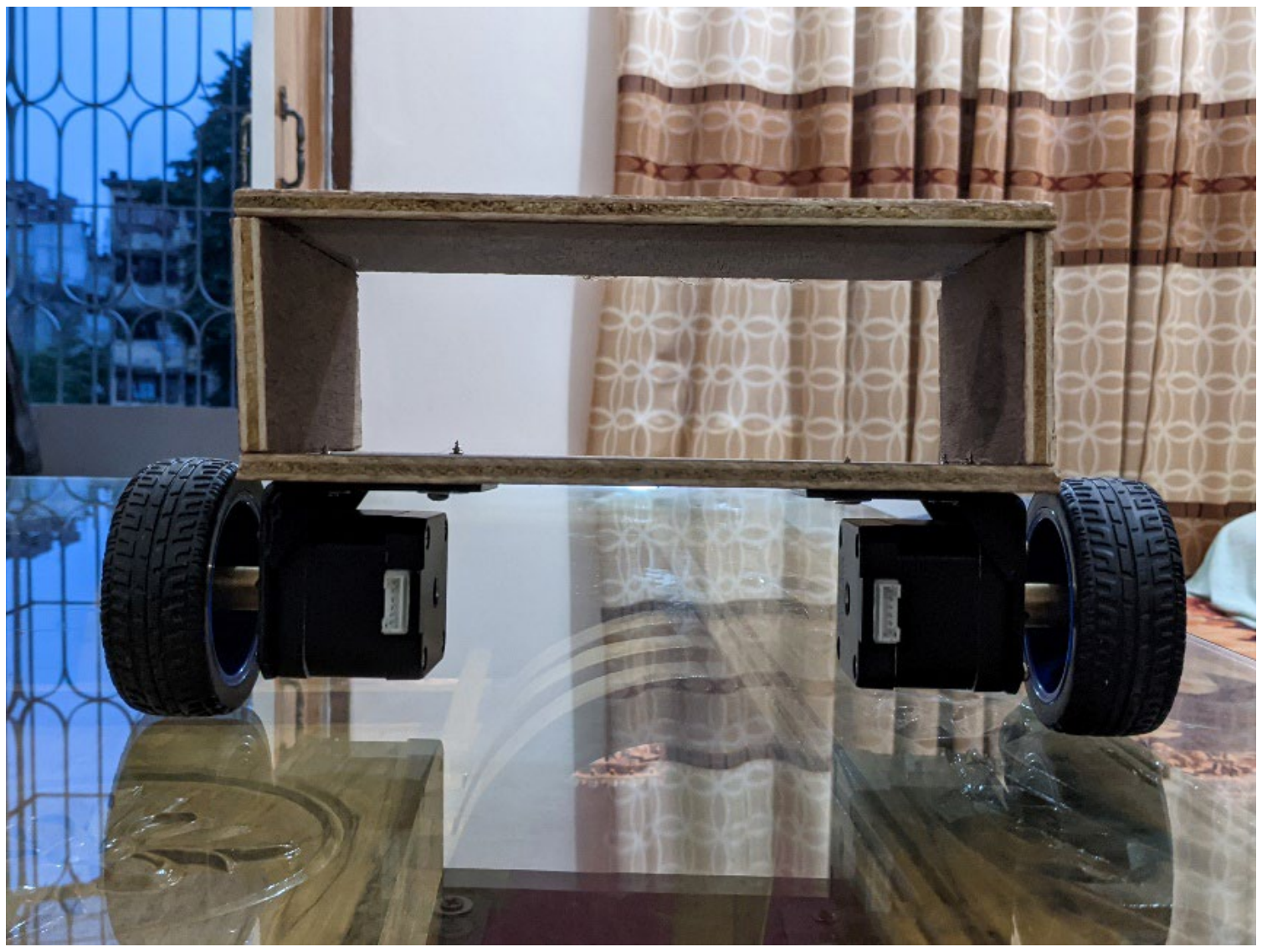

2. Motor brackets: The used stepper motor being heavy, they required special brackets to join to the body of the bot.

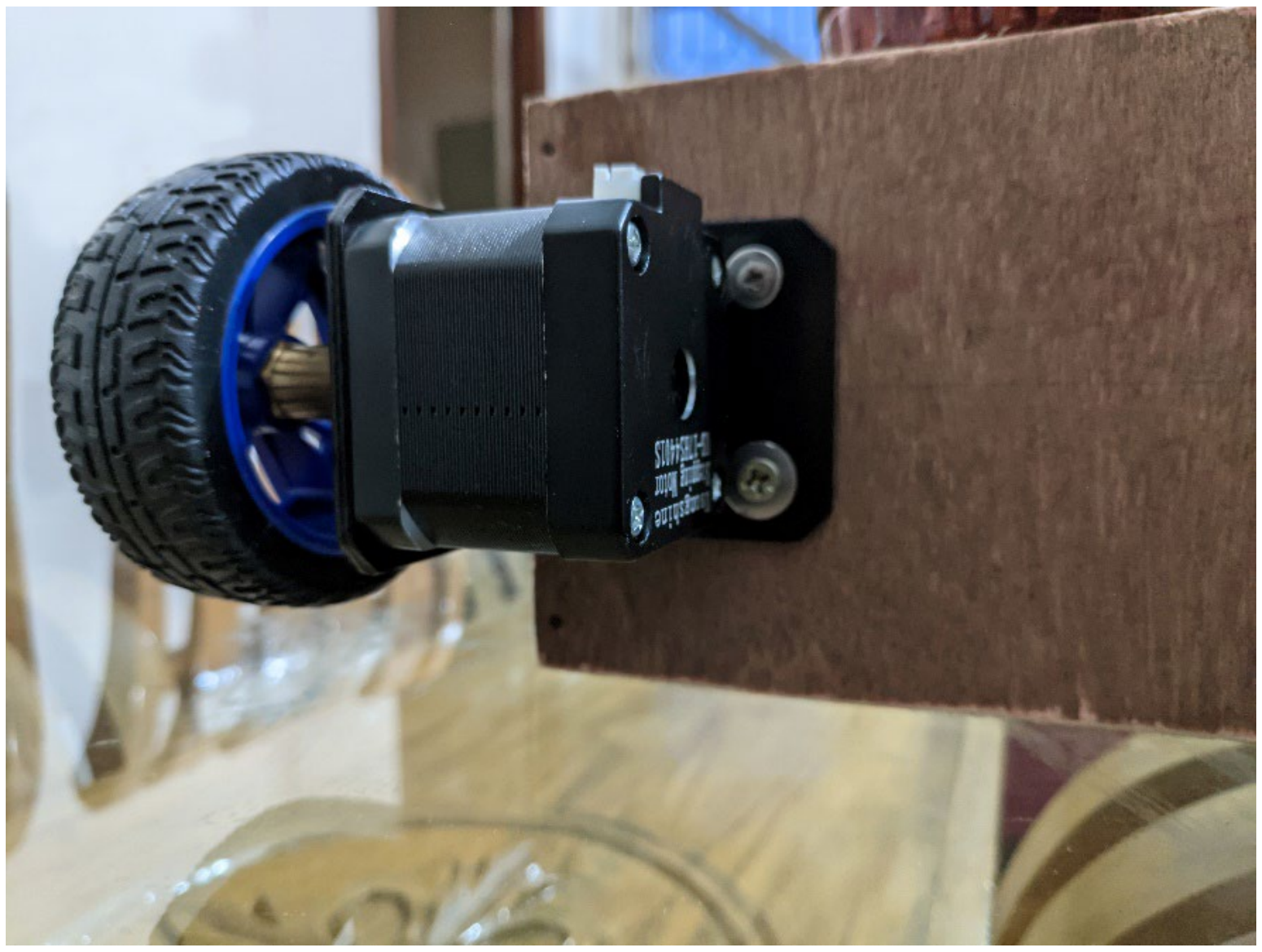

3. Wheels: Plastic wheels with rubber exterior were used for better grip. The wheel were 60 mm in diameter and 27mm thick

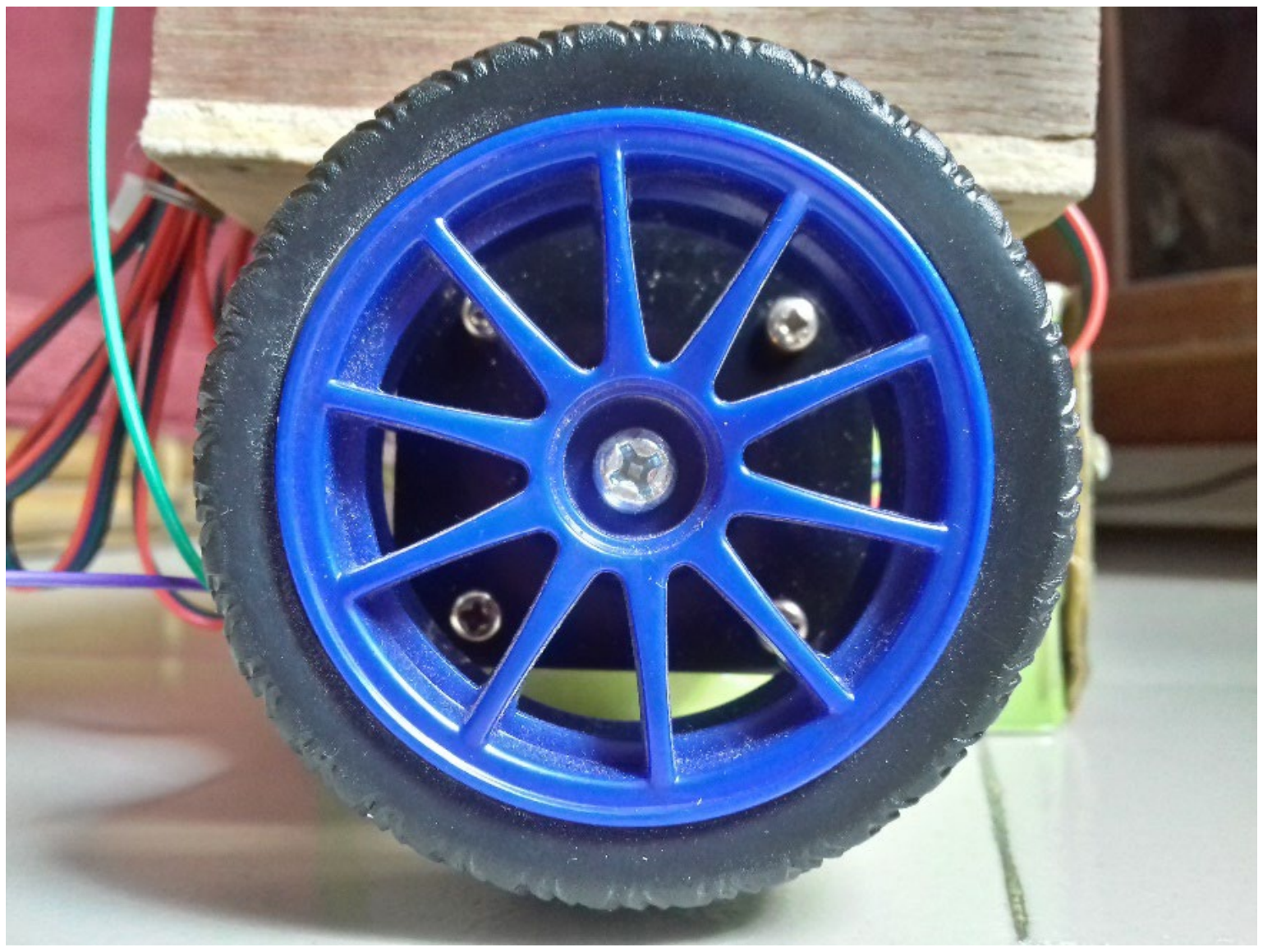

4. Shaft for joining wheel: Auxiliary shafts were needed to join the wheels to the rotor nema17 stepper motor

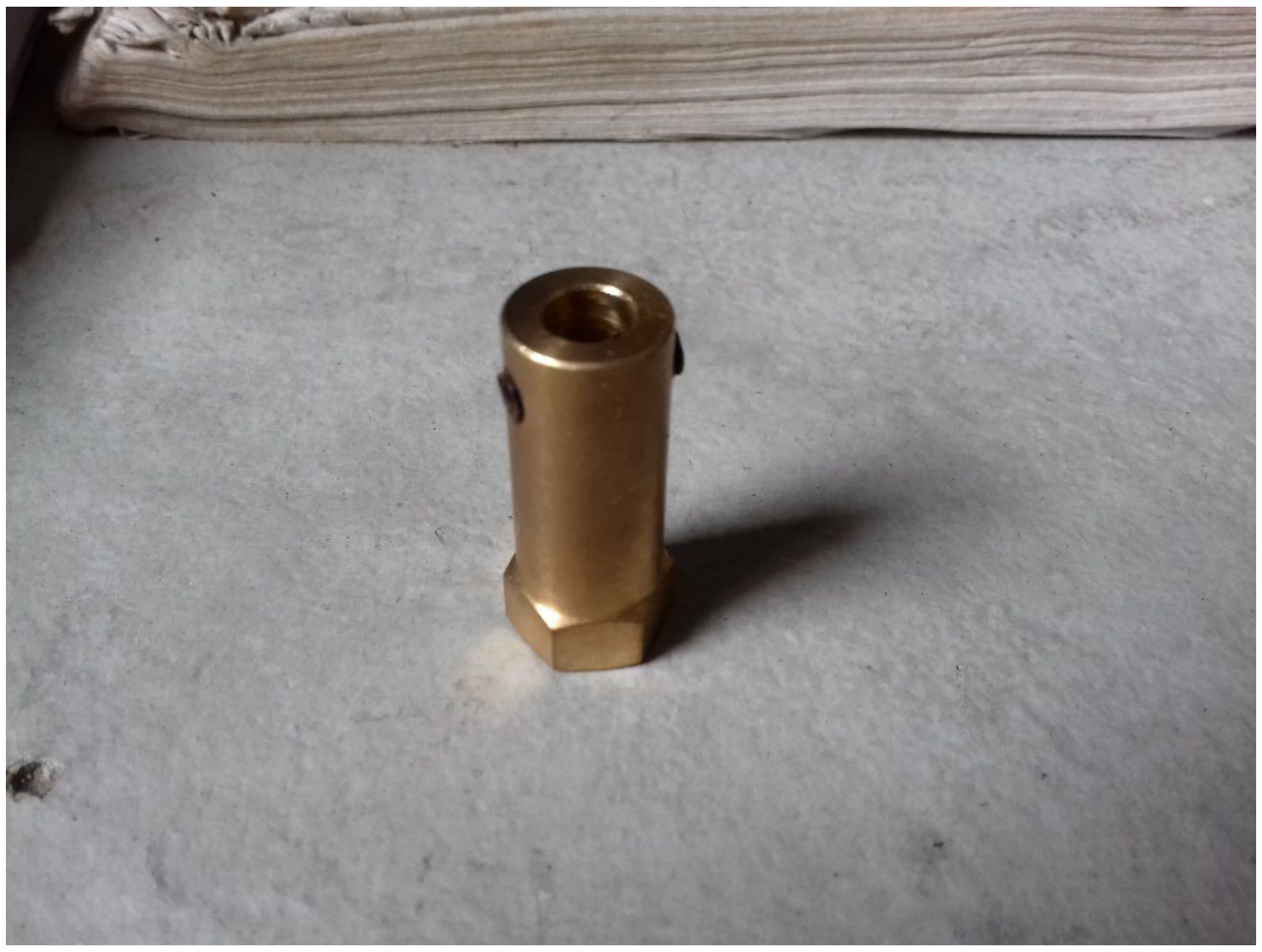

**4.2 Electrical:**

1.Arduino Nano (microcontroller board): The Arduino Nano is a small, complete, and breadboard-friendly board based on the ATmega328. The Arduino has 14 digital in-/output pins and 6 analogue pins. It operates with a clock speed of 16 MHz and supports I2C communication used with the accelerometer and gyro module. The following is the picture of a Arduino nano.

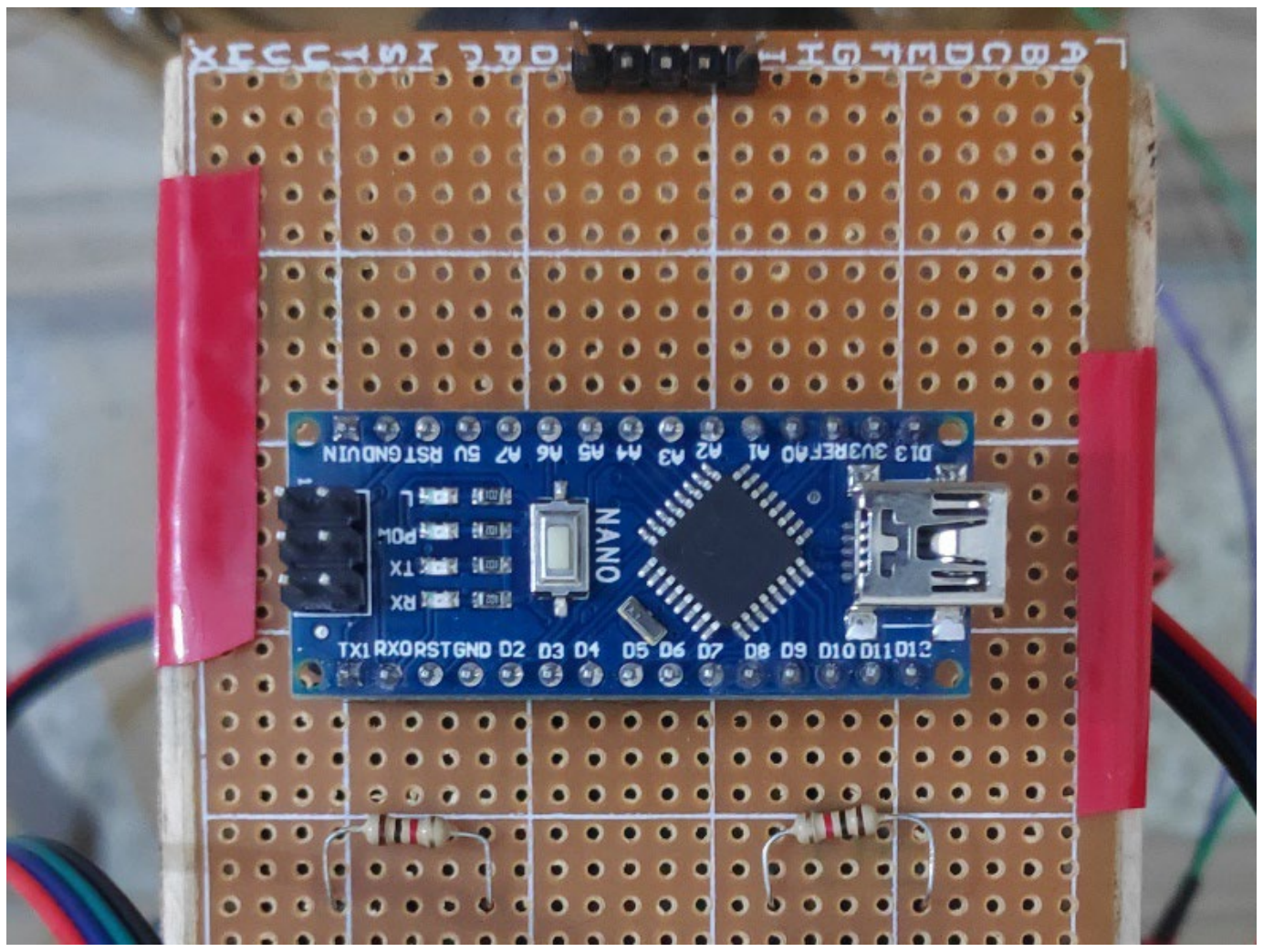

2.MPU 6050 (IMU): To determine the angular deviation and angular velocity an IMU was used. the MPU6050 is a three-axis accelerometer and gyro. It communicates with the Arduino via I2C at 400 KHz, the range for the angular velocity was set to $\pm 250^{\circ}$/s and the range for the accelerometer was set to ±4g.

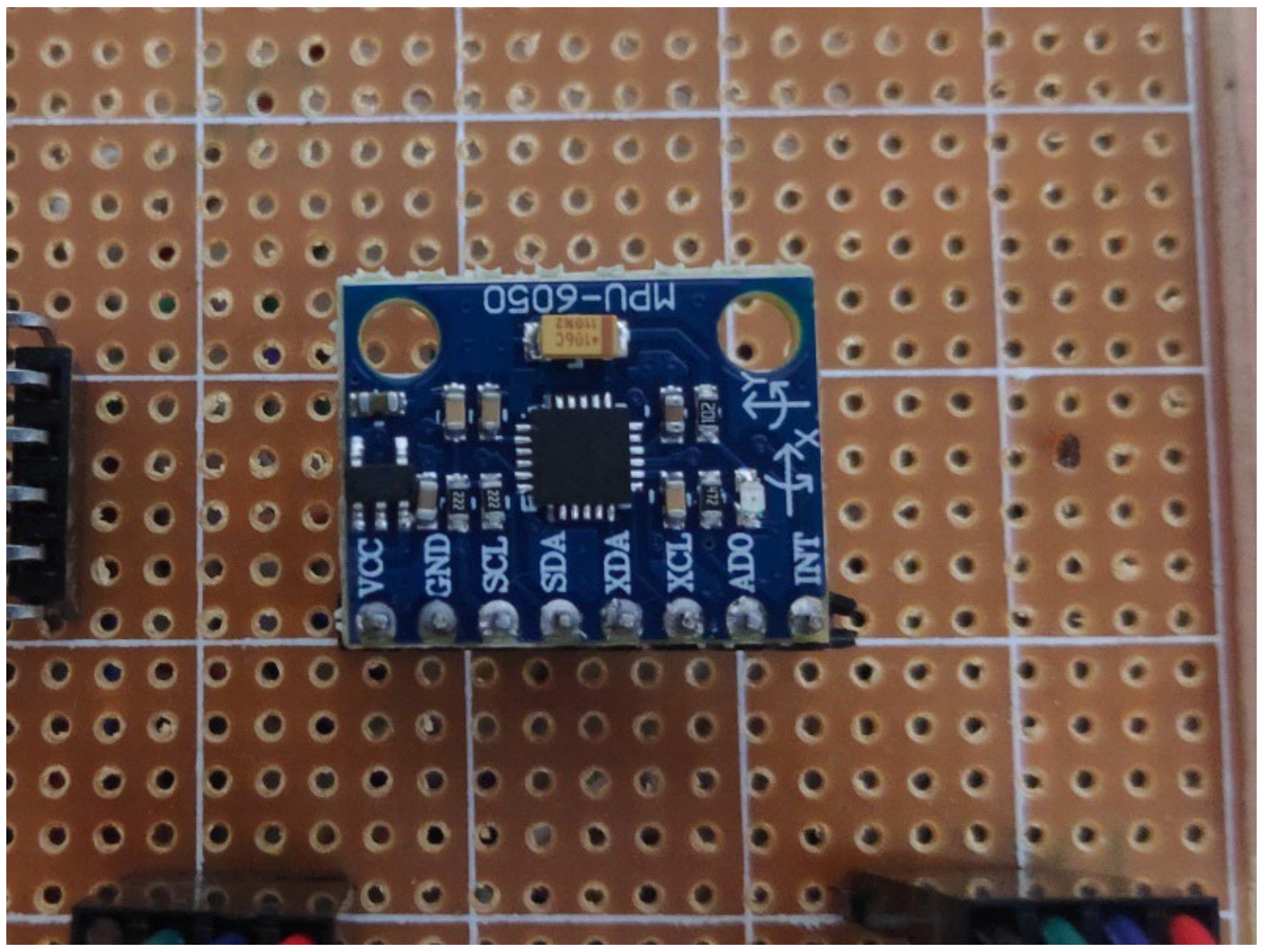

3. A4988 (Motor driver): This breakout board for Allegro’s A4988 micro stepping bipolar stepper motor driver features adjustable current limiting, overcurrent, and over-temperature protection, and five different micro step resolutions (down to 1/16-step). It operates from 8 V to 35 V and can deliver up to approximately 1A per phase.

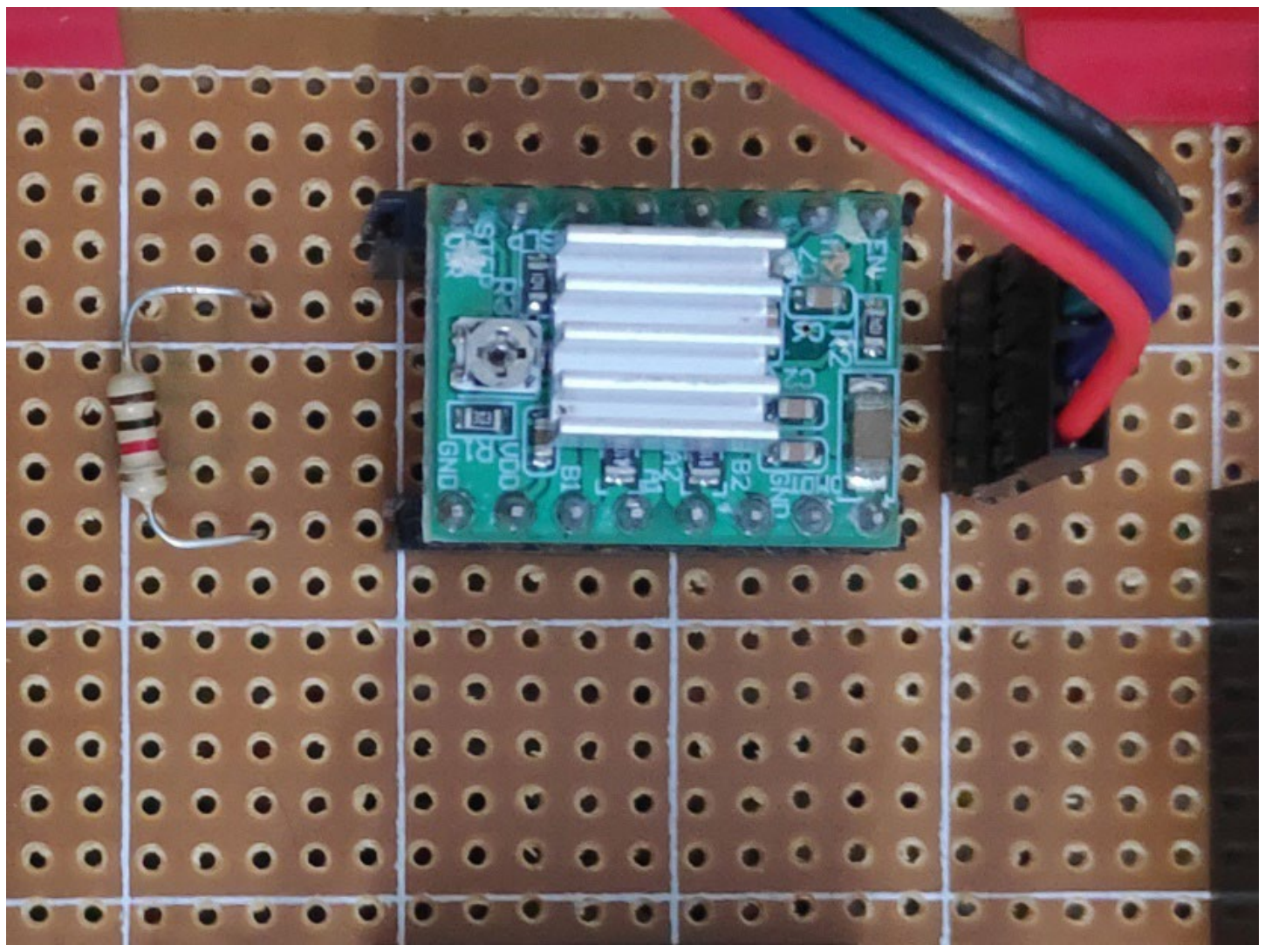

4. Stepper motor (NEMA 17): NEMA 17 is a hybrid stepping motor with a 1.8° step angle (200 steps/revolution). Each phase draws 1.2 A at 4 V, allowing for a holding torque of 3.2 kg-cm.

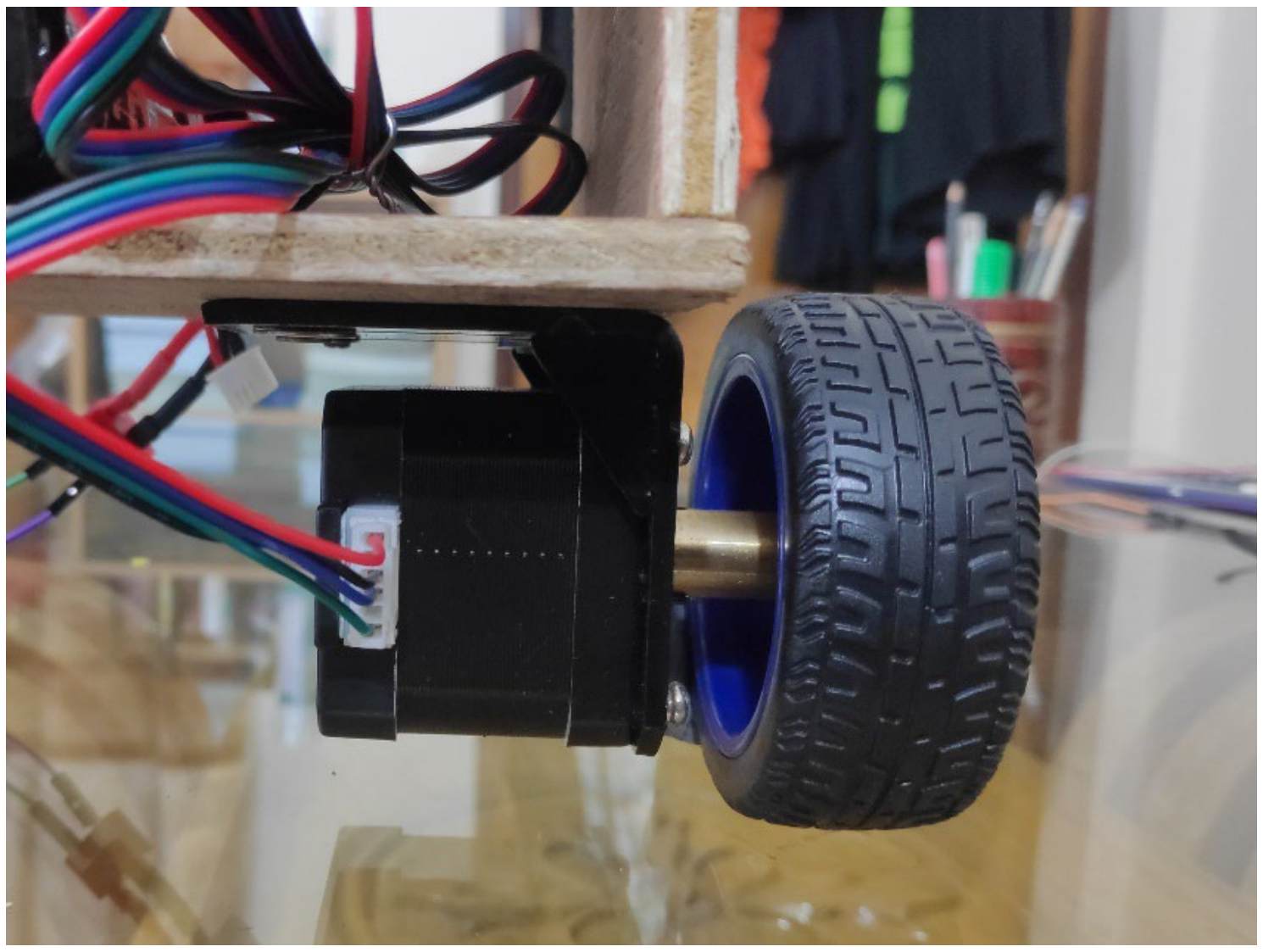

5. HC-05 Bluetooth module: Designed to replace cable connections HC-05 uses serial communication to communicate with the electronics. Usually, it is used to connect small devices like mobile phones using a short-range wireless connection to exchange files. It uses the 2.45GHz frequency band.

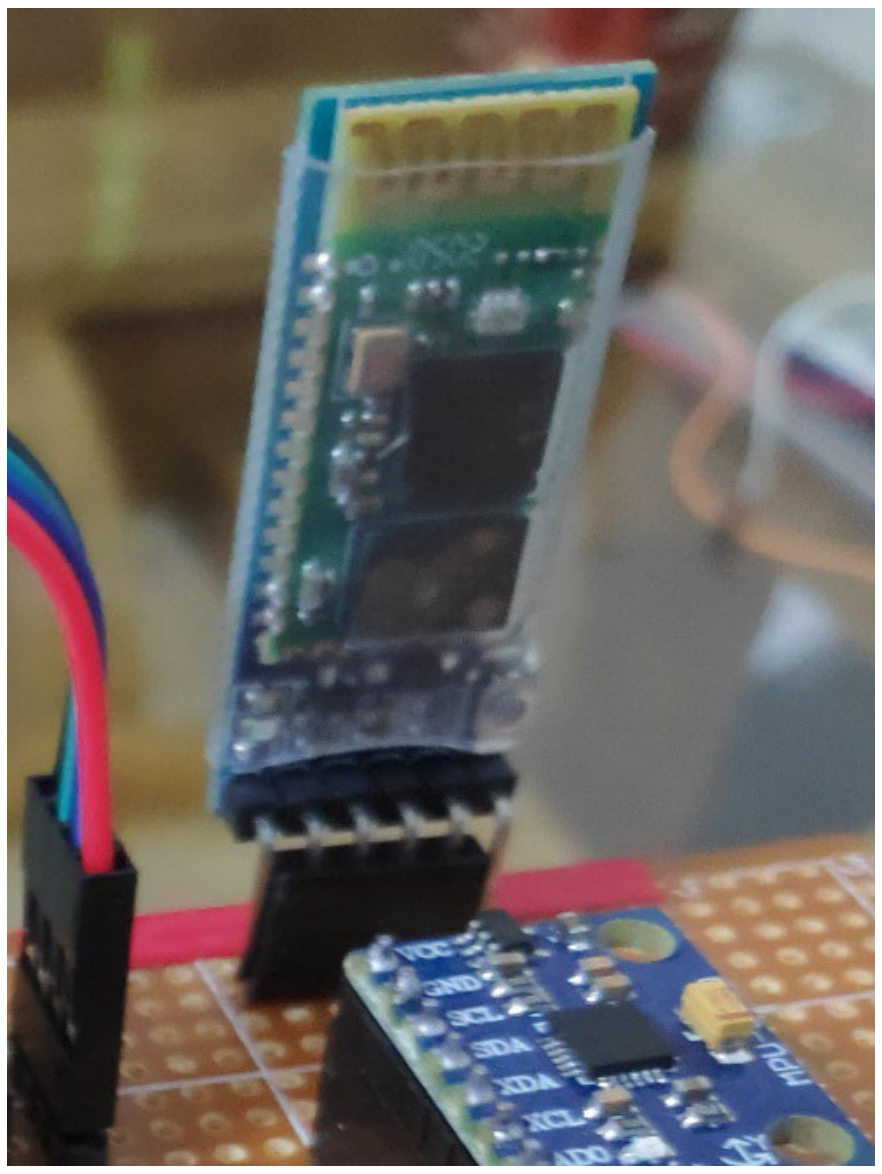

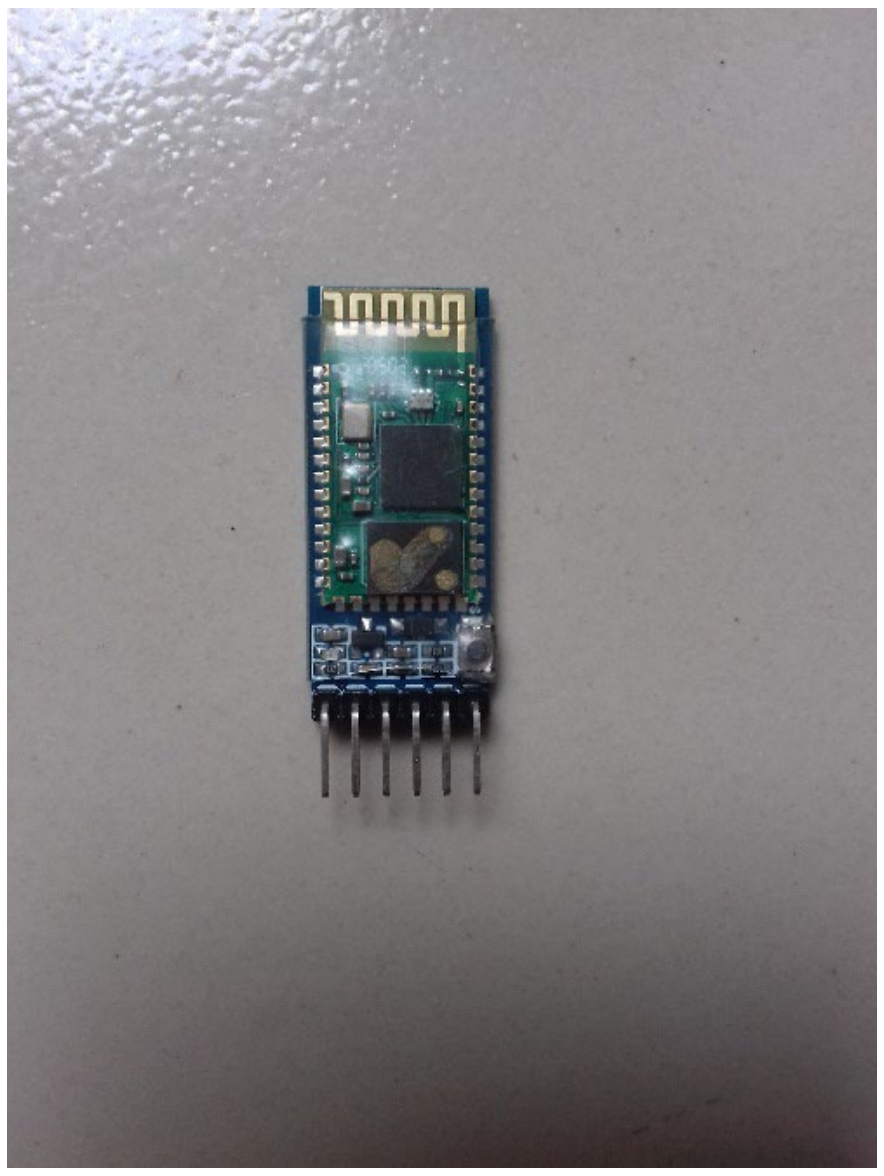

6. Battery: A 11.1V 1100mAh LiPo Battery was use as our main power supplier

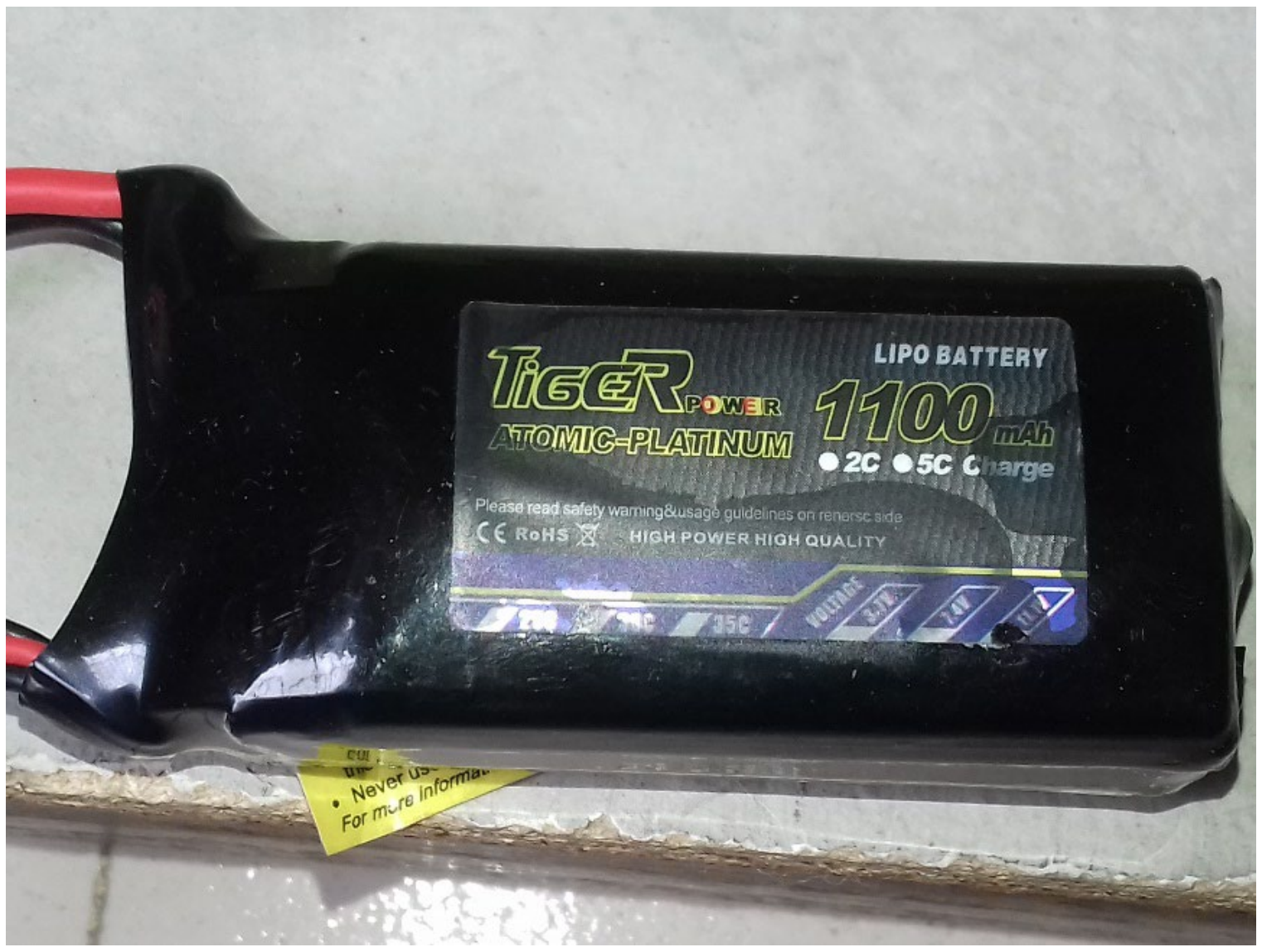

**4.3 Finished Assembly:**

The components were slotted on to a Veroboard via female headers. The header pins were soldered onto the board with necessary wirings among them. This design choice was made to have the opportunity to swap out faulty components easily. Also, to protect them from possible damage due to soldering. The connections made on the Veroboard follows the circuit diagram below:

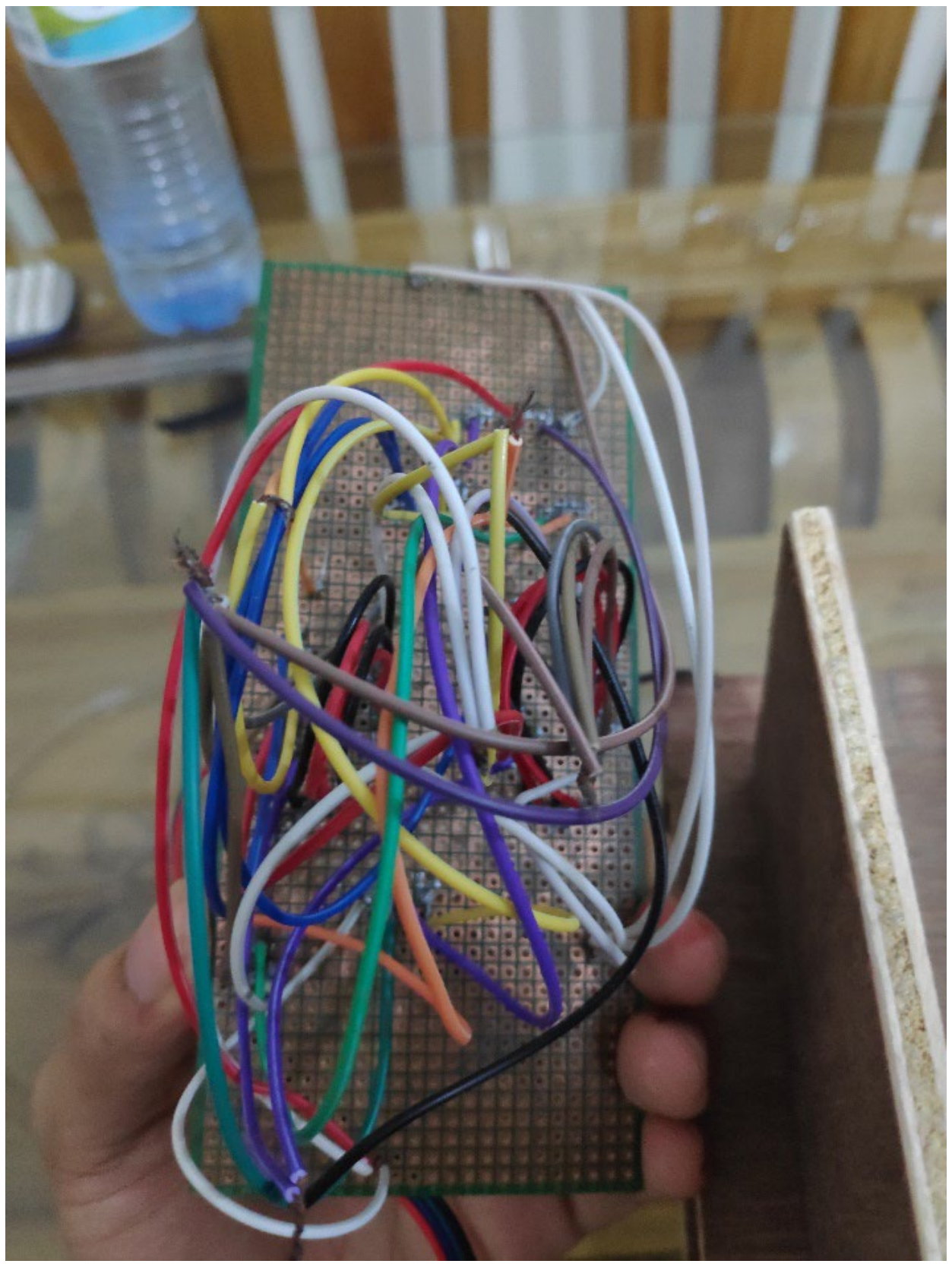

This circuit board was Placed on the housing made on the bot's body. The stepper motor was attached to the bottom of the bot by using brackets. They are connected to the board via cables provided with them.

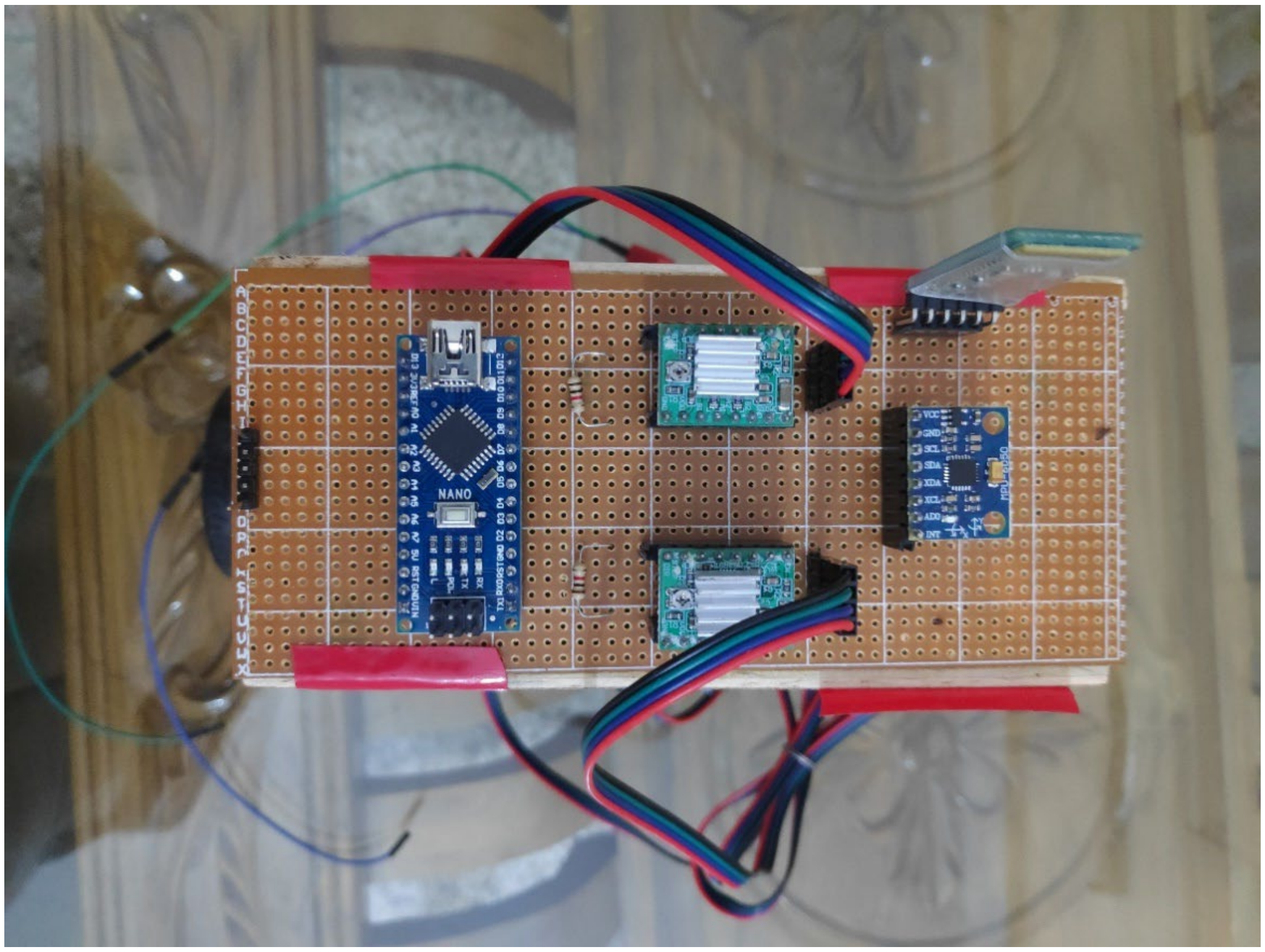

Here is a look at the final assembled robot.

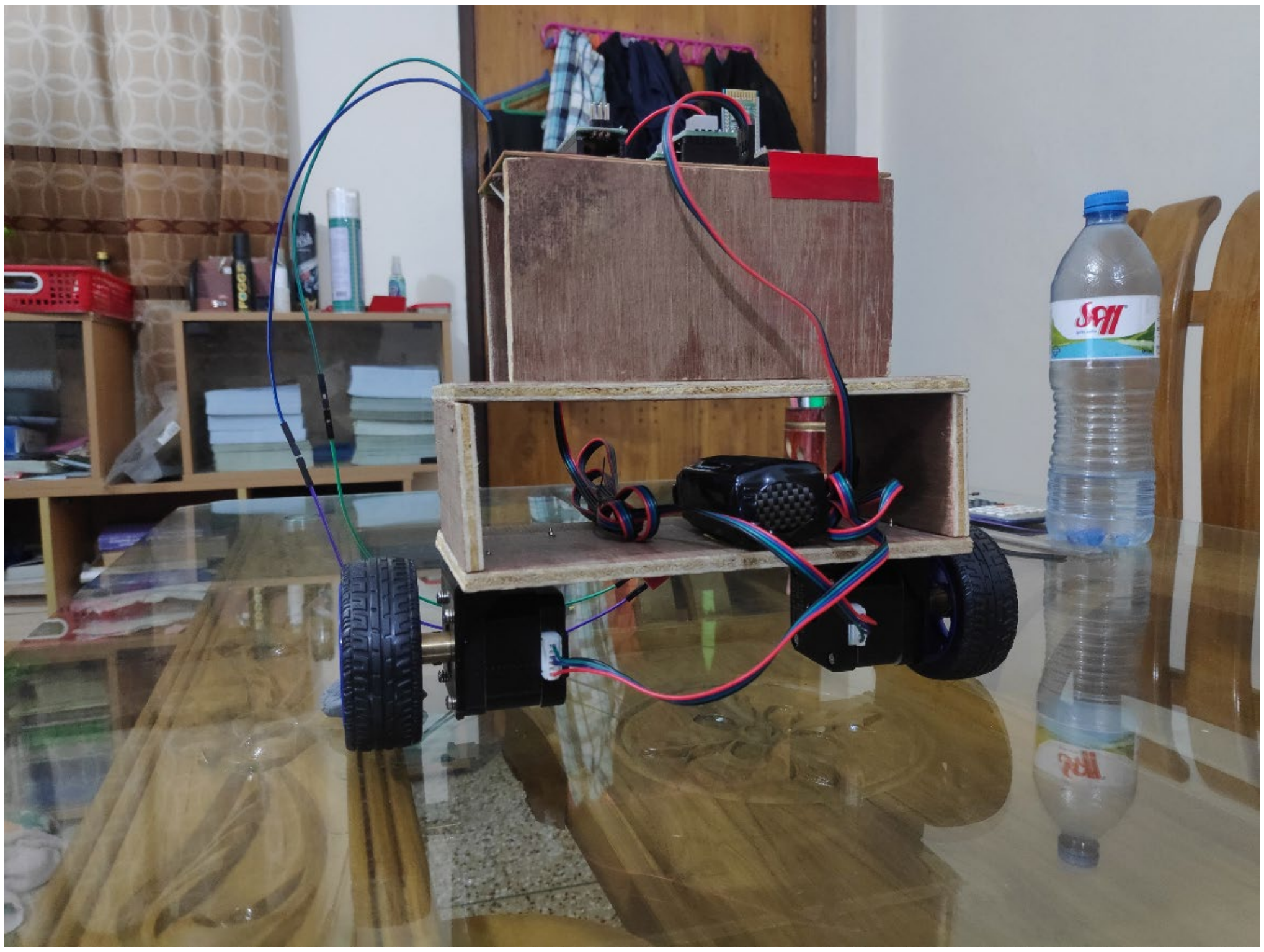

# 5. Calculation, Results and Discussion:

Unfortunately, the robot failed to fulfill some of its objectives due to the sensors not working. When the robot was put together the monitoring software in the controller app was not registering any data from the sensor. When it was taken apart and the sensors tested, it was found out that the sensor is not reading any data. The output in Arduino ide serial monitor is attached here:

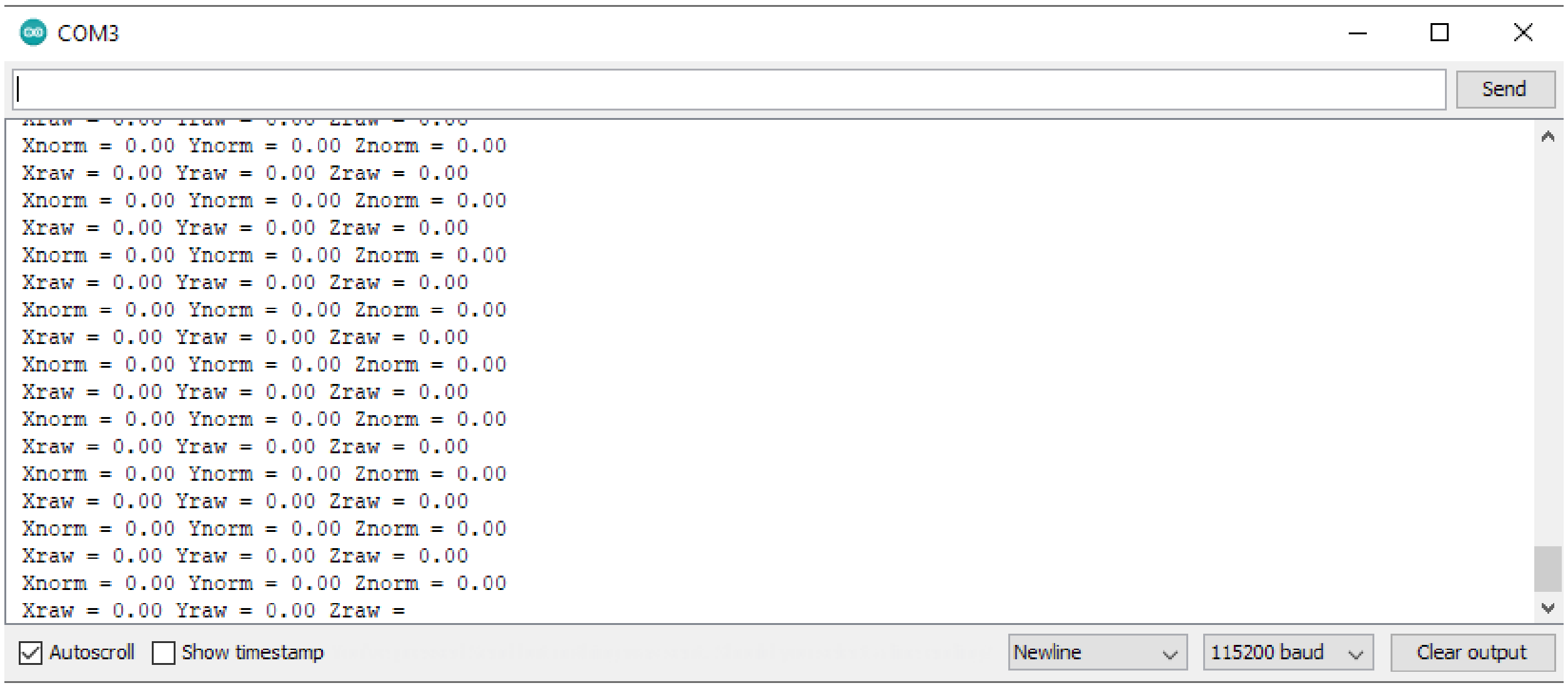

Despite of changing orientation the sensors outputs were not changing. Another sensor was ordered but in the current COVID situation it reached late. And the sensors received were found faulty again and there was not enough time to order and receive another sensor. Without a working sensor the whole system couldn't perform as it supposed to be.

# 6. Conclusions:

Our main objective of this course was to build a self-balancing system which can be operated remotely. The mechanical design of the project, electrical connections, and necessary coding project, all of it was ready to be put together for the completion of the project. But getting two faulty sensors in row was not something we predicted in our initial planning. With no time at hand, there was no other way to devise an alternative method to mimic the sensor's function. The feedback loop of the control system of our bot is dependent on the gyro sensor. With such a crucial component of our system missing, the project is partially completed for the demonstration. But we have plans to continue the work and go beyond the initial goals for our project.

# 7. Reflections on Learning:

Since parts of our project were distributed among our group members, testing and checking different stages of work was not possible. So, when the pieces of work were put together, we had too many problems to troubleshoot and too little time at hand. So, in future we would be aware to keep physical components in one place to do testing and troubleshooting. Also, electronic components being unreliable, it should be made a practice to have spare parts in advance even when a component is working, for unforeseeable scenarios.

## 8. Future work:

Had the project been completed, our next plan regarding this project would have been implementing the self-balancing robot in our day-to-day life. Carrier bots are being used any many places. Designing the bots with a self-balancing system will allow them to be more compact and move in limited environments. This system can be used for locomotion of most types of robots. So, this project has the potential for widespread uses. To advance further, it is possible to reduce one more wheel and use one omni-directional wheel to maneuver this bot. This will allow the bot to move from any position to any direction. Besides the project is interesting enough to introduce newcomers to the field of control engineering and its implementation.